\documentclass{article}

\title{Analysis and Modeling of 3D Indoor Scenes\footnote{Technical Report, School of Computing Science, Simon Fraser University, SFU-CMPT TR 2017-55-3}}
\author{Rui Ma}
\date{\vspace{-4ex}}

\usepackage{amsmath,amsfonts,amssymb,amsthm}
\usepackage[pdfborder={0 0 0}]{hyperref}
\usepackage{graphicx}
\usepackage{caption}

\usepackage{url}
\newcommand{\Rmnum}[1]{\uppercase\expandafter{\romannumeral #1}}

\begin{document}

	\maketitle
	\begin{abstract}
	We live in a 3D world, performing activities and interacting with objects in the indoor environments everyday. 
	Indoor scenes are the most familiar and essential environments in everyone's life.
	In the virtual world, 3D indoor scenes are also ubiquitous in 3D games and interior design.
	With the fast development of VR/AR devices and the emerging applications, the demand of realistic 3D indoor scenes keeps growing rapidly.
	Currently, designing detailed 3D indoor scenes requires proficient 3D designing and modeling skills and is often time-consuming.
	For novice users, creating realistic and complex 3D indoor scenes is even more difficult and challenging.
	
	Many efforts have been made in different research communities, e.g. computer graphics, vision and robotics, to capture, analyze and generate the 3D indoor data.
	This report mainly focuses on the recent research progress in graphics on geometry, structure and semantic {\em analysis} of 3D indoor data and different {\em modeling} techniques for creating plausible and realistic indoor scenes.
	We first review works on understanding and semantic modeling of scenes from captured 3D data of the real world.
	Then, we focus on the virtual scenes composed of 3D CAD models and study methods for 3D scene analysis and processing.
	After that, we survey various modeling paradigms for creating 3D indoor scenes 
	and investigate human-centric scene analysis and modeling, which bridge indoor scene studies of graphics, vision and robotics.
	%
	%
	At last, we discuss open problems in indoor scene processing that might bring interests to graphics and all related communities.  
\end{abstract}
	\section{Introduction}



\begin{figure}[b] \centering
    \includegraphics[width=\linewidth]{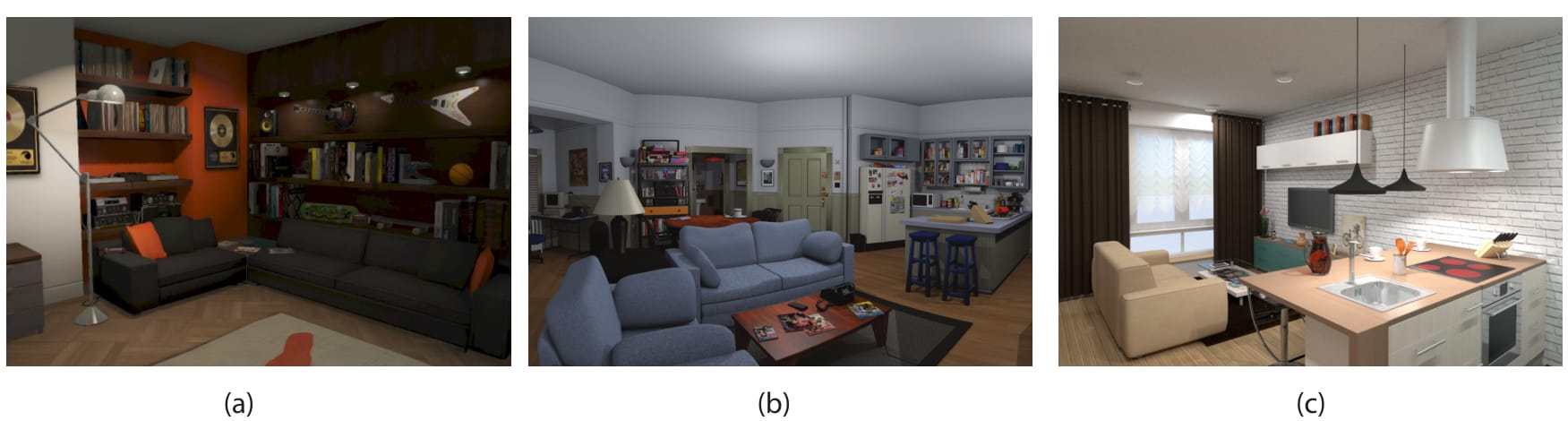}
    \caption{Examples of 3D indoor scenes: (a) a living room from game \textit{GTA \Rmnum{5} \copyright{Rockstar Games}}; (b) \textit{Jerry's Place} modeled for virtual reality~\cite{vr_jerry}; (c) a studio created for interior design~\cite{planner5d}.}
    \label{fig:scenes_eg}
\end{figure}

We live in a 3D world, performing activities and interacting with objects in the indoor environments everyday. 
Indoor scenes are the most familiar and essential environments in everyone's life.
In the virtual world, indoor scenes are also ubiquitous in 3D games, interior design and various VR/AR applications, e.g. VR therapy, military and medical training, crime scene investigation~\cite{vr_wiki} (see Figure~\ref{fig:scenes_eg} for some examples).
Especially, with the fast development of head-mounted display (HMD) devices like {\em Oculus, HTC Vive} and {\em Google Cardboard}, virtual reality is becoming more and more popular in recent years.
To create a better virtual experience, realistic 3D indoor scenes are needed to provide living environments for the virtual characters.

Currently, most existing 3D indoor scenes in online repositories are manually created by general 3D modeling softwares such as 3DS Max and SketchUp, or professional indoor scene modeling tools, e.g., Autodesk Homestyler~\cite{homestyler}, Sweet Home 3D~\cite{swhome}, Planner 5D~\cite{planner5d}.
Creating a scene using these tools often requires proficient 3D designing and modeling skills,
while designing \emph{detailed} and \emph{complex} 3D indoor scenes requires even more expertise and is very time-consuming.

In the research community, many efforts have been made for 3D indoor scene processing, from semantic modeling of indoor environments using captured data, analysis of existing 3D scenes, to synthesizing novel scenes by data-driven systems.
The topics cover problems in a wide range of domains, e.g., computer graphics, vision and robotics.
In this report, we mainly focus on the recent research progress in computer graphics on geometry, structure and semantic {\em analysis} of 3D indoor data and different {\em modeling} techniques for creating plausible and realistic indoor scenes.
In each topic, we explore the application of the data-driven approach which has been widely studied for shape analysis~\cite{xu2017}, 
in context of indoor scene processing. 

\begin{figure}[!t] \centering
    \includegraphics[width=\linewidth]{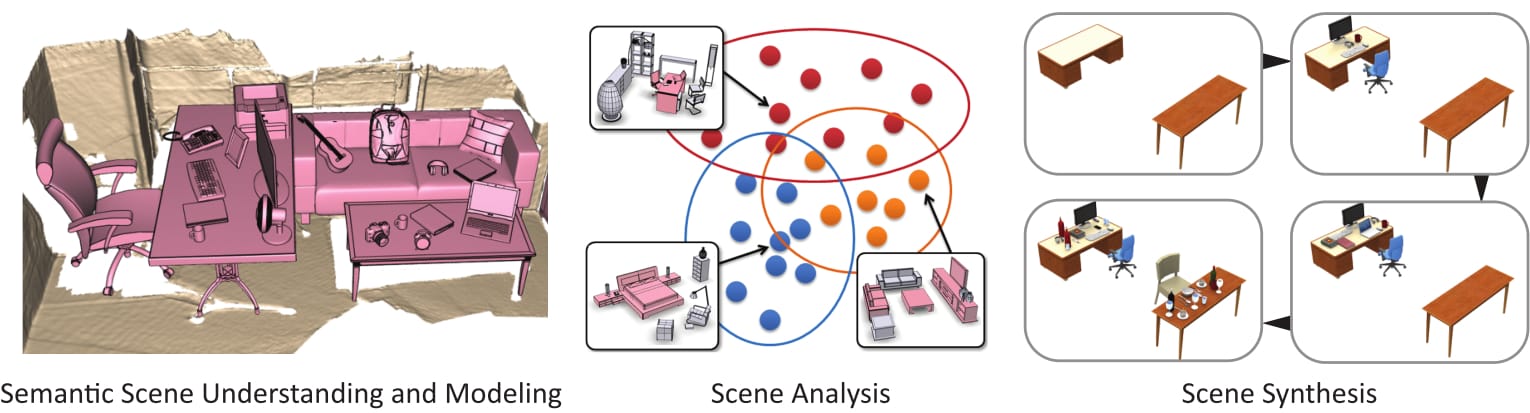}
    \caption{Topics covered in this report.}
    \label{fig:scene_topics}
\end{figure}

One way for generating indoor scenes is to capture the real world indoor environments and convert them into 3D scenes.
There has been a great deal of work in computer vision on 3D indoor scene \emph{reconstruction}~\cite{seitz2006, knapitsch2017},
which takes as input one or more images or depth scans and aims to reproduce the captured scene geometry.
Recent works in graphics~\cite{nan2012, kim2012, shao2012, chen2014, li2015, xu2016} are taking a data-driven and model-based perspective to further \emph{understand} the semantics of the reconstructed data and produce a \emph{semantic modeling} of the scene by replacing the recognized objects with 3D CAD models from the databases.

In addition to reconstructing scenes from real world data, re-using existing scenes from the scene databases, e.g., 3D Warehouse~\cite{3dwarehouse}, is also a way to create and synthesize novel and complex 3D indoor scenes.
As the scene databases usually contain a large number of scenes, efficient scene retrieval and database exploration techniques~\cite{fisher2011, xu2014} are proposed to help a user quickly find the scenes that may be useful.
Moreover, to generate plausible scenes, object occurrences and relationships must be satisfied, e.g., a monitor is normally placed on a desk, and such knowledge could also be extracted from the databases.
\emph{Scene analysis} aims to process the existing 3D scenes in the databases and learn the knowledge and constraints that could be used for arranging objects in new scenes.

With prior knowledge learned from scene databases and other sources, data-driven scene modeling often starts from some user provided inspirations, which could be text~\cite{coyne2001, seversky2006, chang2014b}, sketches~\cite{shin2007, lee2008, xu2013} and images~\cite{liu2015,izadinia2017}, and generates 3D scenes based on the input.
To achieve the goal of obtaining complex indoor scenes, 3D exemplars are also used for scene \emph{synthesis}~\cite{fisher2012}. 
Recently, human-centric scene modeling which has been studied in vision and robotics, is introduced to graphics for functional scene synthesis~\cite{fisher2015} and progressive scene \emph{evolution} by human actions~\cite{ma2016}.

In the rest of this survey, we first review works on understanding and semantic modeling of scenes from captured 3D data of the real world (Figure~\ref{fig:scene_topics} left).
Then, we focus on the virtual scenes composed of 3D CAD models and study methods for 3D scene analysis and processing (Figure~\ref{fig:scene_topics} middle).
After that, we survey various modeling paradigms for creating 3D indoor scenes 
and investigate human-centric scene analysis and modeling (Figure~\ref{fig:scene_topics} right).
At last, we discuss open problems in indoor scene processing that might bring interests to graphics and all related communities.  

	\section{Understanding and Semantic Modeling from Captured 3D Data}
\label{chap:scan}

Creating 3D indoor scenes as the digital replica of our living environments is a common interest of computer vision and graphics.
There has been a great deal of work on 3D indoor scene \emph{reconstruction} which takes as input one or more images or depth scans and aims to reproduce the captured scene geometry~\cite{seitz2006, knapitsch2017}.
Although real-time reconstruction with fine-grained details has been achieved for large scale environments~\cite{newcombe2011, niessner2013},
the reconstructed scene is usually represented as a single 3D model without semantic object segmentation.

Object recognition from captured 3D data is the main task for 3D scene \emph{understanding}.
Given some depth scans or a reconstruction of an indoor environment, scene understanding is to identify and segment the objects from the background,
classify them into semantic categories with prior knowledge learned from database or cluster similar objects into groups in the unsupervised case.
One direct application of scene understanding is \emph{semantic modeling}, which converts the scanned scene into a 3D scene composed of 3D CAD models.
In the modeling process, for each recognized object, a 3D model is retrieved from the model database based on the object's semantic category and geometry features observed from the corresponding object segmentation.
Then the retrieved models are transformed and placed into the scene, yielding a full 3D scene that has the same semantic objects and the same arrangements as the input capture.

In this chapter, we survey recent works that use model-based method as well as unsupervised approach for 3D scene understanding and semantic modeling.
Then we study online scene modeling which identifies objects during the real-time scene reconstruction process.

\subsection{Model-based Scene Understanding and Semantic Modeling}
Missing data, scanning noise and lack of prior knowledge place significant challenges for 3D scene understanding.
Since indoor scenes often contain a large number of objects which are arranged in a complex manner,
scans of cluttered scenes may have large portion of missing data due to the object occlusion, e.g., chairs pulled underneath the desks.
Also, some objects or their parts are difficult to capture, e.g., small items on the desk, legs of table and object parts made with glass.
Noises and missing data are produced because of the scanner's resolution, specular reflection from the environment lighting, and the thin structure of object's geometry.
Even with a well-captured scene, object recognition may still be difficult as indoor objects usually have large geometrically and pose variations, and different deformation and articulations, e.g., the book and laptop are opened or closed.
To address these problems, recent works take a data-driven and model-based approach which learns the prior knowledge of objects from the scene databases or 3D models and applies learned knowledge for semantic scene understanding and modeling.

\begin{figure}[b] \centering
    \includegraphics[width=\linewidth]{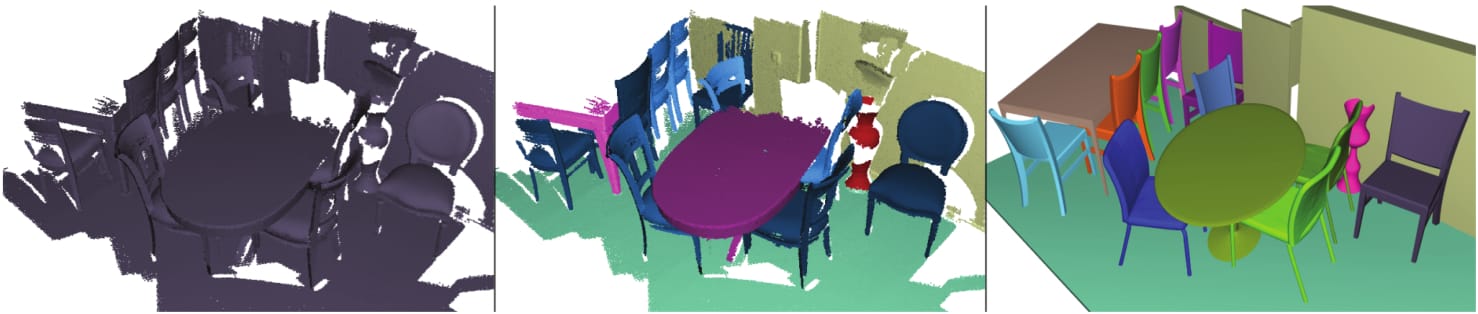}
    \caption{Indoor scene understanding by Search-Classify~\cite{nan2012}. Given a raw scan of a highly cluttered scene (left), the search-classify method is applied to segment the scene into meaningful objects (middle), followed by a template based deform-to-fit reconstruction (right).}
    \label{fig:nan2012}
\end{figure}

Nan et al.~\cite{nan2012} propose a search and classify approach for object recognition and reconstruction of scanned 3D indoor scenes (Figure~\ref{fig:nan2012}).
In this work, object classification and segmentation is argued as a \emph{chicken-egg} problem, that object classification cannot be directly applied to the scene since object segmentation is unavailable, while object segmentation is also challenging as the spatial relationships between points and patches are neither complete nor reliable.
Their key idea to solve this problem is to interleave object segmentation and classification in an iterative manner.

The system first trains a classifier from a large set of 3D object scans for recognition of several specific object classes, e.g., chair, table, cabinet, monitor.
Given a raw scan of highly cluttered scene, the algorithm first over-segments the scene into piecewise smooth patches.
Then, a search-classify process is performed by iteratively accumulating patches that form regions with high classification likelihood, until the whole scene is classified into meaningful objects and outliers.
To solve the ambiguous cases in classification and remove the outliers, a \emph{deform-to-fit} step is applied to reinforce the classification by deforming templates to fit the classified point cloud and selecting the best matching template.
The output is a consistent scene understanding and an approximate reconstruction which conveys the general object arrangement and inter-relationships as the input scan.

The work of Kim et al.~\cite{kim2012} consider both the low quality of data and object articulation when acquiring and understanding large indoor environments.
With focus on interiors of office buildings or auditoriums, this paper addresses the problem of object recognition by exploiting three observations: indoor objects in such environments often come from a small number of categories (e.g., wall, floor, chair, table, computer) and repeat many times; these objects usually consist of rigid parts, while the variability and articulation are limited and low-dimensional;
mutual relationships among the basic objects satisfy strong priors, e.g., a monitor rests on the table.

A pipeline with two phases is presented:
in a learning phrase, each object of interest is scanned for a few times (typically 5 - 10 scans across different poses) to construct primitive-based 3D models and recover their modes of variation;
in a recognition phase, from a single-view scan, objects with different poses are segmented and classified based on the low complexity models generated in the learning phase as well as the inter- and intra-object relationships.
By requiring additional acquisition of more object models and their variability modes in the learning phase, the pipeline could be extended to the understanding and modeling of more complex indoor environments.

\begin{figure}[!t] \centering
    \includegraphics[width=\linewidth]{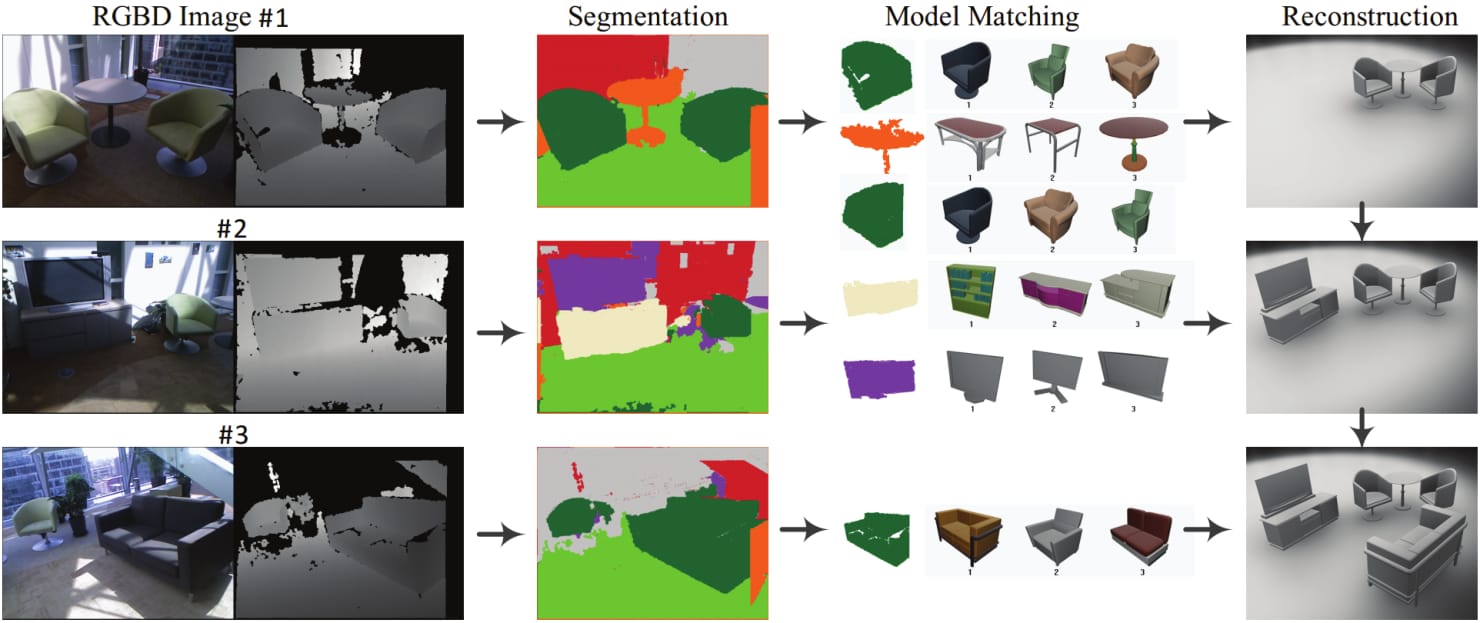}
    \caption{Pipeline of interactive semantic scene modeling~\cite{shao2012}.}
    \label{fig:shao2012}
\end{figure}

Targeting at semantic modeling of indoor scenes with a consumer-level RGB-D camera,
Shao et al.~\cite{shao2012} present an interactive system that takes one or a few sparsely captured RGB-D images and produce a 3D scene with semantic objects and their layouts matched to the input (Figure~\ref{fig:shao2012}).
First, the user captured RGB-D indoor images are automatically segmented into a set of regions with semantic labels.
If the segmentation is not satisfactory, the user can draw some strokes to guide the algorithm to achieve better results.
After the segmentation is finished, the depth data of each semantic region is used to retrieve a matching 3D model from a database.
Each model is then transformed according to the image depth to yield the scene.
By using multiple images sharing sufficient overlaps between successively captures, medium or large scale scenes can be modeled by unifying 3D models built for each image.

To reduce the user interaction in~\cite{shao2012}, Chen et al.~\cite{chen2014} exploit contextual relationships learned from a 3D scene database to assist the object segmentation and classification for automatic semantic modeling.
As indoor objects normally have strong contextual relationships, e.g., object co-occurrences and spatial arrangements, using context could help remove uncertainty due to noise and missing data by seeking semantically compatible models when scene understanding.
In~\cite{chen2014}, contextual relationships are first extracted from a 3D scene database collected from 3D Warehouse.
Then a context-based two-layer classifier is trained and used for scene modeling: 
the first-layer maps point clouds to semantic objects and their parts and the second layer finds the best matched 3D model with the assigned semantic label.
To refine the modeling result, object contours extracted from the RGB images of the input are further used to recover the small objects and missing parts.

Using knowledge learned from model or scene databases, model-based scene understanding is able to recognize objects from low-quality captures of cluttered indoor scenes and produce a semantic scene modeling with only one or a few views of the environment.
While current model-based methods only focus on a few object categories that frequently appear in the indoor environments,
with more data and a more complicated learning stage, 3D scenes with more complexity could be modeled from the real world.

\subsection{Unsupervised Scene Understanding and Modeling}
Unlike the model-based methods which require a training phase to learn the 3D geometry of interested objects,
unsupervised methods~\cite{karpathy2013, mattausch2014, shao2014} have been studied for scene understanding and modeling recently.
Without the knowledge extracted from the database, such methods use other cues, e.g., the intrinsic geometry properties and the repetition of indoor objects, or the physical stability of object arrangements, to identify the objects from the indoor scans.

Karpathy et al.~\cite{karpathy2013} present a method for discovering objects from reconstructed 3D meshes of indoor environments.
The key observation is certain geometric properties are useful in discovering indoor objects, even when no semantic label is provided.
Given a scene reconstruction using existing tools, e.g., KinectFusion~\cite{newcombe2011},
the algorithm first decomposes the scene into a set of candidate mesh segments and then ranks each segment according to its \emph{objectness} -- a quality defined based on the intrinsic shape measures (e.g., symmetry, smoothness, local and global convexity).
The output is a set of ranked object hypotheses which could be used for suggestion of object bounding boxes and reduce the time-consuming object labeling process in 3D scenes.

\begin{figure}[t] \centering
    \includegraphics[width=\linewidth]{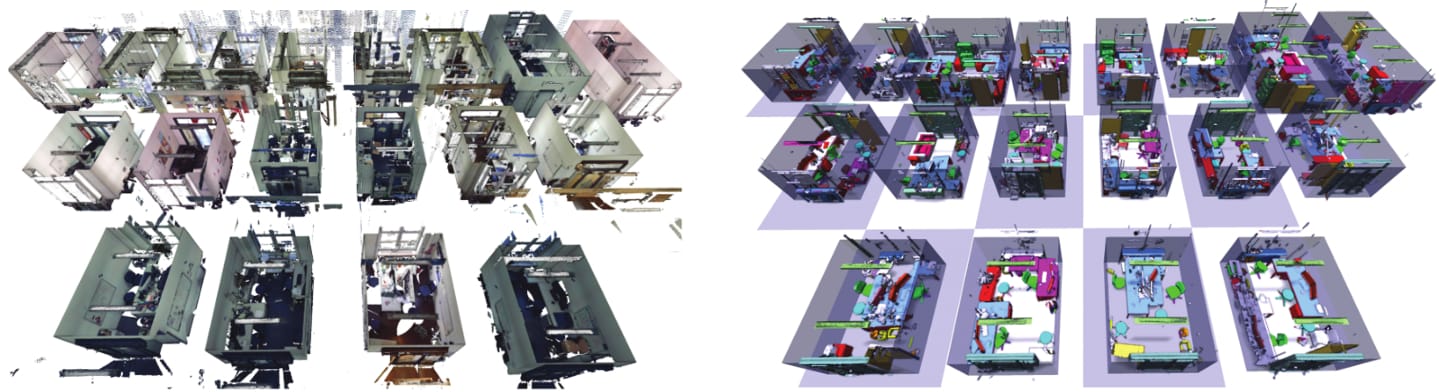}
    \caption{Unsupervised object detection by~\cite{mattausch2014}. From a set of indoor scans (left), objects (shown in color on the right) are extracted by clustering the planar patches fitted to the point clouds.}
    \label{fig:matt2014}
\end{figure}
Repetition is another cue for unsupervised object detection.
Mattausch et al.~\cite{mattausch2014} present an algorithm to automatically segment a set of indoor scenes by detecting repeated objects (Figure~\ref{fig:matt2014}).
The key idea is to use the similarities within scenes to segment scans of building interiors into clusters of similar objects, and segment individual objects within each cluster.
Point clouds of the input scans are first converted into a collection of planar patches.
A patch similarity measure based on shape descriptors and spatial configuration of neighboring patches is proposed and used for patch clustering.
In each scene, objects are extracted by considering the proximity as well as the separation between the clustered patches.
Only using planar patch for clustering, it is difficult for~\cite{mattausch2014} to model small objects with many planar regions, like keyboards or desk lamps.

\begin{figure}[t!] \centering
    \includegraphics[width=\linewidth]{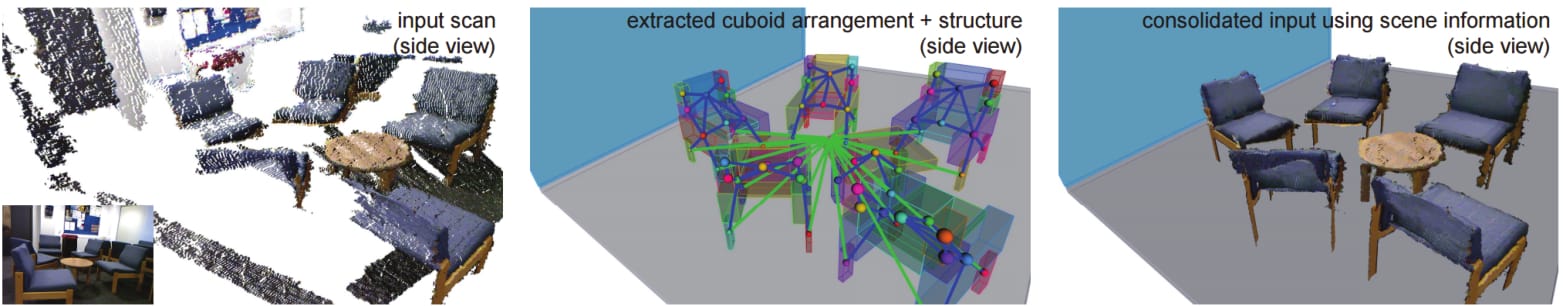}
    \caption{Stability-based scene understanding in~\cite{shao2014}. Starting from a heavily occluded single view RGB-D image (left), a coarse scene structure as an arrangement of cuboids along with their inter-cuboid relations (middle) is extracted using physical stability analysis, and then used for scene understanding and completion (right).}
    \label{fig:shao2014}
\end{figure}

Besides geometric features, analysis of physical stability is explored in Shao et al.~\cite{shao2014} for scene understanding (Figure~\ref{fig:shao2014}).
To address the problem of missing data from the incomplete scans, this work abstracts indoor environments as collections of cuboids (oriented boxes) and hallucinates geometry in the occluded regions by globally analyzing the physical stability of the resultant arrangements of the cuboid.
The algorithm starts by creating a set of initial cuboids from an incomplete 3D point cloud:
the points are clustered using a connected component analysis based on spatial proximity, and fitted into a set of cuboids in a greedy manner.
Next, the initial cuboids are extended to the occluded regions with consideration of the physical stability of the arrangements, which is measured similarly as in~\cite{whiting2009}.
The covered cuboid-based structure is then used for structure-guided completion by registering a good-quality scan of an object to the partial scans.
Semantic modeling without relying on the semantic object labels is also enabled by retrieving the database models using the cuboid structures.

Unsupervised methods provide a new perspective for scene understanding and semantic modeling.
However, as the key assumptions, e.g., planar patches, cuboid structures are only applicable for a limited set of indoor objects,
current works can only handle a few indoor scene types.
Possible solutions to this problem include studying more completed shape features as in~\cite{karpathy2013} or involving human or robot interactions for scene understanding~\cite{zhang2015,xu2015}.

\subsection{Online Scene Reconstruction with Object Recognition}

Recent works have investigated scene understanding and modeling during the process of online scene scanning.
Similar to previous works which perform offline analysis~\cite{nan2012, kim2012, shao2012, chen2014, mattausch2014}, object retrieval from shape databases~\cite{li2015} and structure analysis~\cite{zhang2015} are introduced in the setting of real-time scene reconstruction.
Moreover, ~\cite{xu2015, xu2016} address the problem of autonomous object-level scene reconstruction by a robot, reliving human from the laborious task of detailed scene scanning.

\begin{figure}[!h] \centering
    \includegraphics[width=0.5\linewidth]{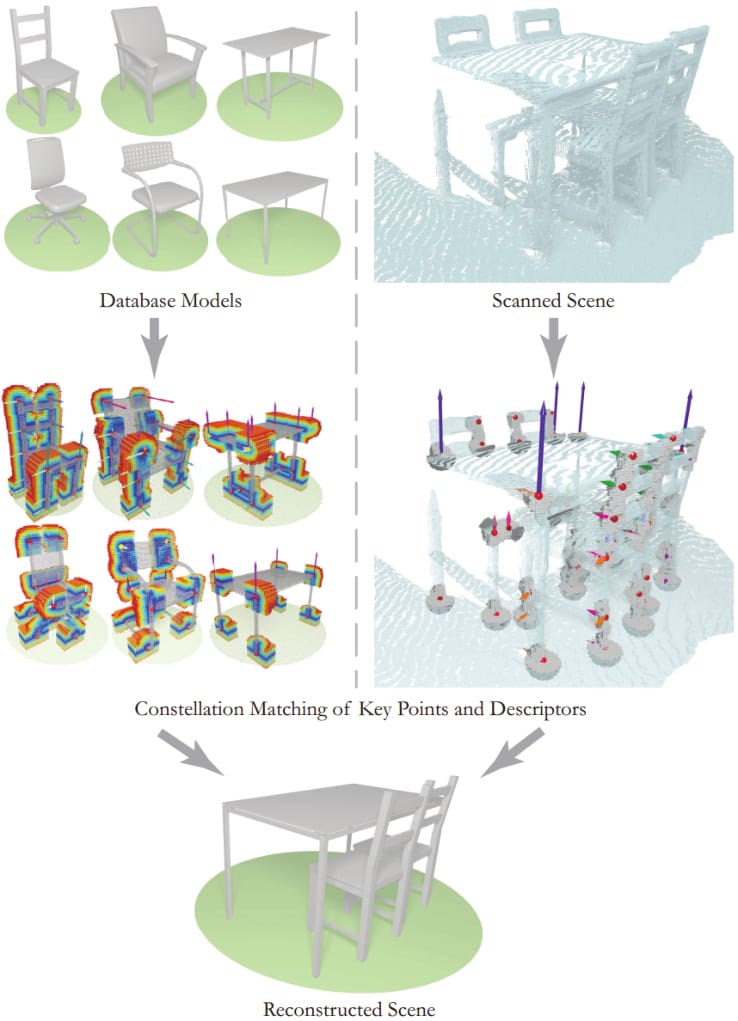}
    \caption{Pipeline of object retrieval for real-time 3D reconstruction~\cite{li2015}.}
    \label{fig:li2015}
\end{figure}
Instead of aiming for reconstructing the accurate scene geometry, Li et al.~\cite{li2015} turn the real-time scene reconstruction into a object retrieval problem (Figure~\ref{fig:li2015}).
The goal is to recognize objects directly from 3D scanning data and replace them with 3D models retrieved from shape databases at runtime to achieve a complete and detailed scene reconstruction.
The main challenge for shape retrieval and registration in the context of 3D scanning is the partial and noisy data requires a robust partial matching that finds a 3D model based on the scan from only a partial view of the object.

To solve this problem, a novel point-based 3D shape descriptor that combines local information and global geometric context is defined. In the pre-processing, each database model is converted into a point cloud and the point descriptors are computed for a set key points which are detected by a combined 3D/2D corner detector.
These descriptors are then arranged into \emph{constellations} of key points and form the shape descriptor for each model.
At runtime, from the real-time volumetric fusion data~\cite{niessner2013}, key points and the corresponding descriptors are computed and then used for searching key point constellations of database models.
The matchings also produce transformations for registering the models to the scene.
In the end, a semantic modeling of the scanned environment is obtained. 

Similar to~\cite{mattausch2014}, cues of planar regions and non-planar objects are utilized in~\cite{zhang2015} for real-time scene reconstruction.
Based on the observation that indoor environments usually consist of many planner surfaces (e.g., floors and walls) and isolated objects (e.g., chairs and computers), Zhang et al.~\cite{zhang2015} perform an online structure analysis on the scanned geometry to extract such planes and objects to form a structured scene.
The input of the system is the captured depth maps from voxel hashing based scene reconstruction~\cite{niessner2013}.
A plane/object labeling step is followed to achieve segmentation of planes and objects.
Through plane-based structure analysis, plane and object geometries are refined to produce a final reconstruction result.
To reduce the time spent for repeat scanning of nearly identical objects, e.g., chairs of different heights,
duplicate objects are also detected during online analysis.
The user is allowed to interact with the system to decide whether to use the mesh of an already extracted object or keep scanning if she wants to preserve the subtle differences between objects.

\begin{figure}[t] \centering
    \includegraphics[width=\linewidth]{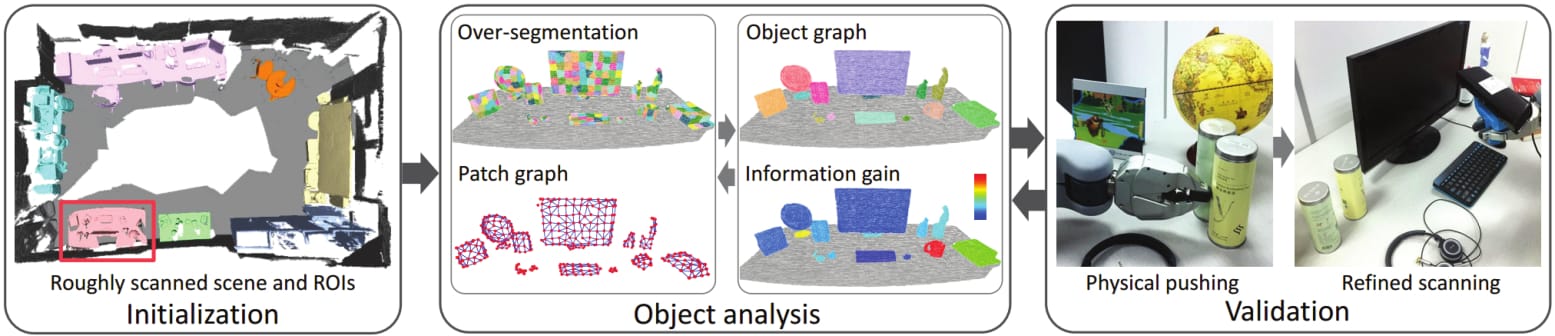}
    \caption{An overview of robot-operated autoscanning for coupled scene reconstruction and object analysis~\cite{xu2015}.}
    \label{fig:xu2015}
\end{figure}
As detailed scene scanning by human is laborious, especially for large indoor scenes containing numerous objects,
Xu et al.~\cite{xu2015} propose a framework for robot-operated autonomous scene scanning or autoscanning (Figure~\ref{fig:xu2015}).
The autoscanning and reconstruction is object-aware, guided by object-centric analysis which incorporates online learning to the task of object segmentation.

The basic system setup is a mobile robot holding a depth camera performing real-time scene reconstruction~\cite{niessner2013}.
With the current reconstruction, the robot executes an iterative \emph{analyze-and-validate} algorithm:
object analysis is performed to segment the scene into a set of hypothetical objects;
then the joint uncertainty in both object-level segmentation and object-wise reconstruction is estimated and used to guide the robot in validating the reconstruction through physical push and scanning refinement.
The gained knowledge about segmentation as well as the newly acquired depth data is incorporated into current reconstruction, 
which would in turn improve the analysis in the next round.
This interleaving process repeats until the overall uncertainty does not decrease significantly.
The output of the system is a reconstructed scene with both object extraction and object-wise geometry fidelity.
Due to the involvement of physical push from the robot, the proposed method is invasive and it does not reconstruct the original scene with exact object positions.
This is arguably acceptable since exact positioning of movable objects is usually quite casual and accidental in the real-life indoor scenes.

\begin{figure}[b] \centering
    \includegraphics[width=\linewidth]{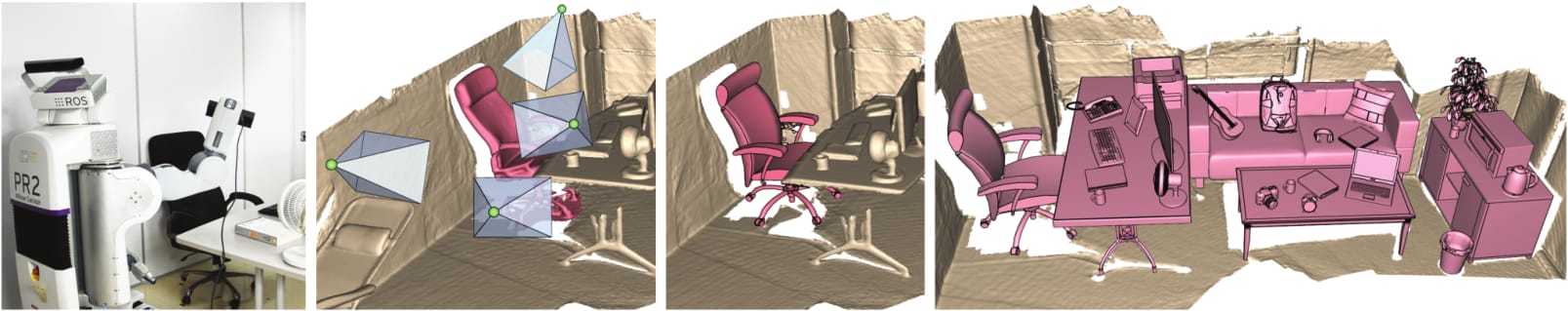}
    \caption{Autonomous active object identification and scene modeling~\cite{xu2016}.}
    \label{fig:xu2016}
\end{figure}
In the follow-up work of~\cite{xu2016}, the problem of autonomously exploring unknown objects in a scene by robot-operated consecutive depth acquisitions is addressed (Figure~\ref{fig:xu2016}).
The goal is to reconstruct the scene while online identifying the objects from a large collection of 3D shapes, with multi-view depth observations.
Autonomous object identification from indoor scans is challenging because of two facts:
on one hand, unlike coarse-level shape classification, e.g., discriminating a chair from a bicycle for which a single view would suffice, fine-grained object identification over a large shape collection requires observations from multiple views;
on the other hand, since the object is unknown, planning the views to best discriminate it with other shapes while considering robot movement cost and recognition efficiency is difficult.

To solve the coupled object recognition and view planning problem, a 3D Attention Model is proposed to select the next-best-views (NBVs) for depth acquisition targeting at an object of interest and conduct part-based recognition to tolerate occlusion.
Given a series of progressively acquired RGB-D images, the task involves first extracting potential object, and then identifying for each object the most similar 3D shape from a database based on the attention model.
The retrieved 3D model is inserted into the scene, to replace the object scan, thus progressively constructing a 3D scene model.


	\section{Analysis and Characterization of 3D Indoor Scenes}

In addition to reconstructing scenes from scanned data of real world environments,
indoor scene creation by modeling with existing 3D scenes is also widely studied in computer graphics community.
3D scenes in current online databases, e.g., 3D Warehouse~\cite{3dwarehouse}, provide a valuable data source for data-driven scene processing.
To generate plausible scenes, 3D scene analysis is often executed to learn the object occurrences and relationships from database scenes as prior knowledge for object placement.
Geometry and structure analysis of 3D scenes also leads to a better understanding of the scene data and enables applications such as scene retrieval~\cite{fisher2011, xu2014} and intelligent scene editing~\cite{sharf2014}.
Furthermore, co-analysis of a collection of scenes allows to efficient organization and exploration of large scene databases~\cite{xu2014}, as well as creation of consistent scene representations~\cite{liu2015} for future tasks.

In this chapter, we first introduce a representative graph-based scene representation which encode structural relationships between objects~\cite{fisher2011}.
Next, we study scene representations that could encode complex object relationships and their applications in 3D scene analysis~\cite{zhao2014, sharf2014}.
In the end, we survey techniques for co-analysis and processing of a collection of scenes~\cite{xu2014, liu2014}.

\subsection{Structural Scene Representation}

\begin{figure}[!t] \centering
    \includegraphics[width=\linewidth]{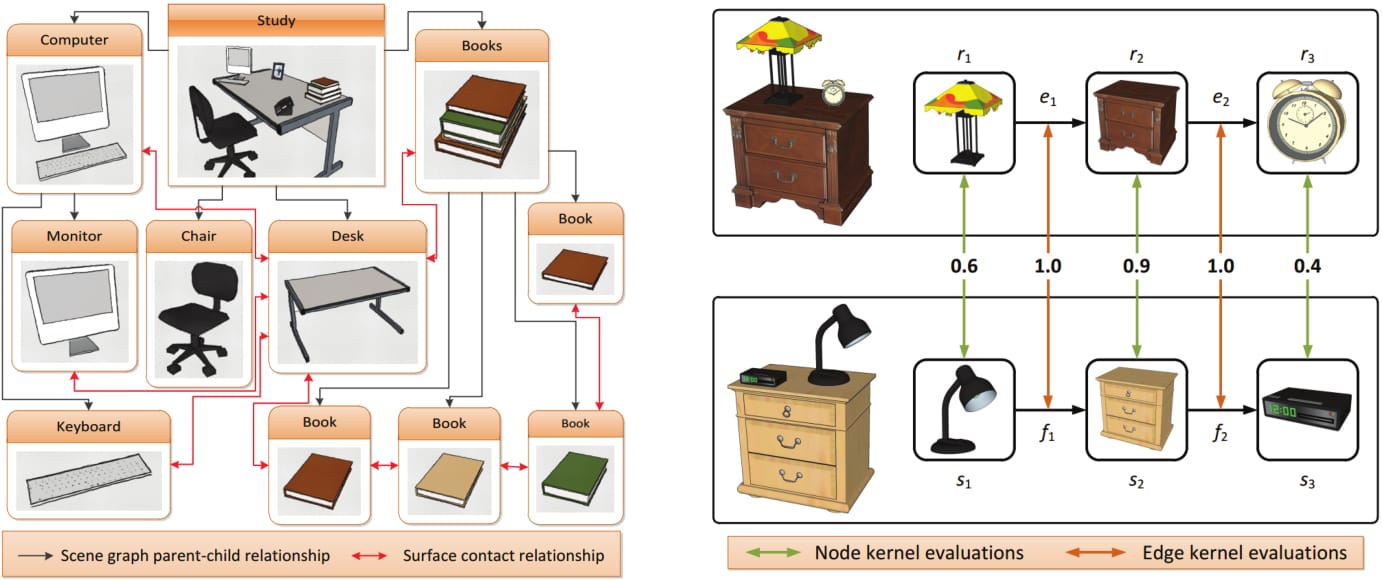}
    \caption{Graph-based scene representation and comparision in Fisher et al.~\cite{fisher2012}: a scene and its representation as a relationship graph (left); comparisons of two graph walks (right).}
    \label{fig:fisher2011}
\end{figure}

The work of Fisher et al.~\cite{fisher2011} takes a 3D scene and transforms it into a \emph{relationship graph}, a representation which encodes structural relationships between objects, such as support, contact or enclosure.
The nodes of a relationship graph correspond to meaningful objects in the scene and the edges represent the relationship between these objects (Figure~\ref{fig:fisher2011} left).
To build the relationship graph, the algorithm starts with an initial scene graph associated with the scene downloaded from 3D Warehouse.
The initial graph nodes are consolidated by human to represent meaningful objects; relationship edges between two objects are created by geometric tests, e.g., whether two meshes of the objects are contacted.
 
Representing scenes as relationship graphs greatly facilitates comparing scenes and parts of scenes.
A graph kernel is defined for comparison of two relationship graphs:
similarities between the graph nodes and edges are computed and accumulated to produce an overall similarity of two graphs (Figure~\ref{fig:fisher2011} right).
By considering the geometry features and semantic labels of objects encoded in the nodes and structural relationships encoded in the edges,
the graph kernel works well for characterizing 3D scenes and largely improves the performance of scene retrieval.

\subsection{Characterization of Complex Relationships}

\begin{figure}[!t] \centering
    \includegraphics[width=\linewidth]{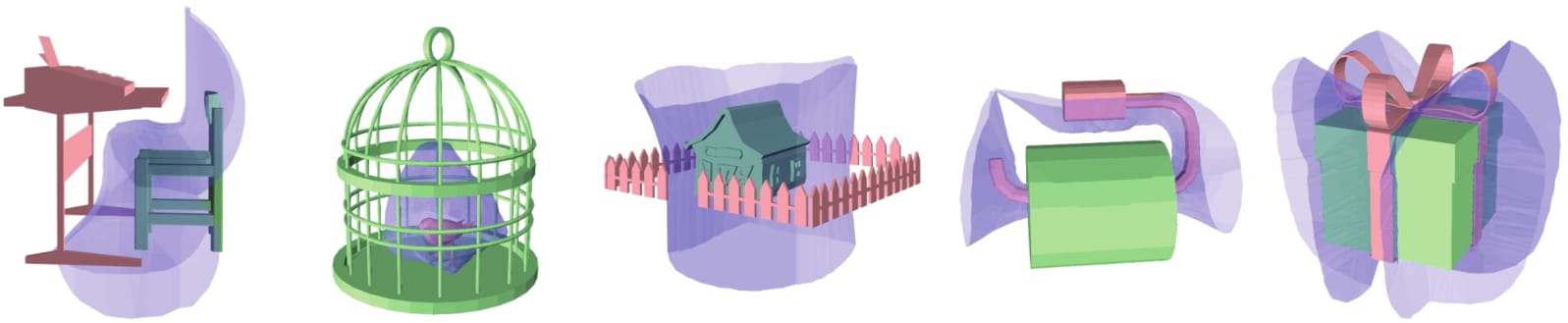}
    \caption{Examples of IBS (in blue) of two objects~\cite{zhao2014}.}
    \label{fig:zhao2014}
\end{figure}

To further represent more complex relationships between objects, Zhao et al.~\cite{zhao2014} propose a new relationship descriptor,
Interaction Bisector Surface (IBS), which describes both topological relationships such as whether an object is wrapped in, linked to or tangled with others, as well as geometric relationships such as the distance between objects (Figure ~\ref{fig:zhao2014}).
Inspired by the application of Voronoi diagram in indexing and recognizing the relationships of proteins in the area of biology,
the IBS is defined as the set of points that are equidistant from two objects, which forms an approximation of the Voronoi diagram for objects in the scene.
By computing a topological feature set called the Betti numbers of the IBS, relationships such as enclosures and windings can be detected.
The geometric relationships are analyzed using the shape of the IBS, the direction of its normal vectors, and the distance between the objects and the IBS.

\begin{figure}[!htbp] \centering
    \includegraphics[width=\linewidth]{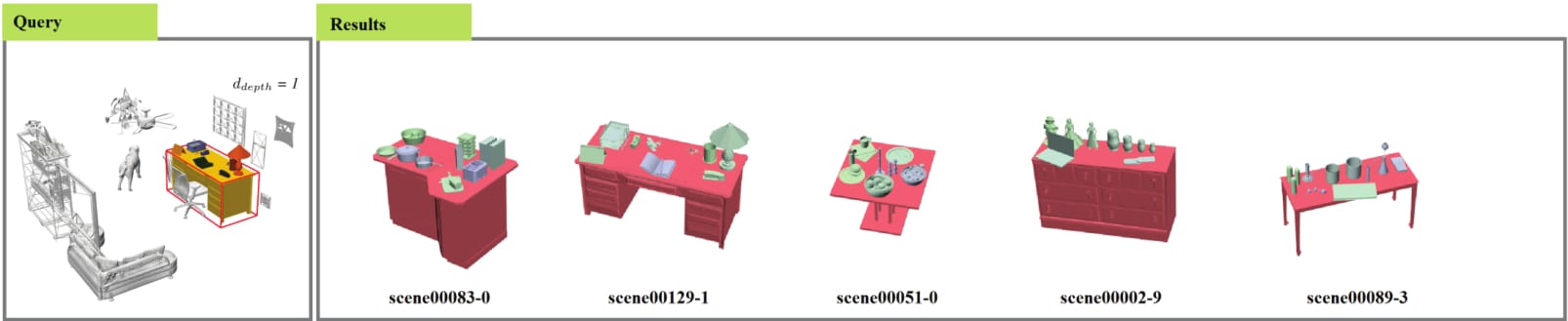}
    \caption{Content-based relationship retrieval based on IBS~\cite{zhao2014}. In the query scene (left), the object with a bounding box is the query object; the resulting scenes (right) are ordered by the similarity and the red object is the retrieved object with a similar context to the query object. }
    \label{fig:zhao2014_ret}
\end{figure}

Automatic construction of scene hierarchies is studied by first using IBS to define a closeness-metric between objects and then grouping individual objects or object groups to form a hierarchy.
As IBS focuses on modeling the interaction between objects, spatial relationships are characterized only based on the topological and geometric features of the IBS, while object labels are not needed as in~\cite{fisher2011}.
Therefore, content-based relationship retrieval is enabled by computing the similarity of interactions based on the IBS features (Figure~\ref{fig:zhao2014_ret}).

\begin{figure}[!htbp] \centering
    \includegraphics[width=\linewidth]{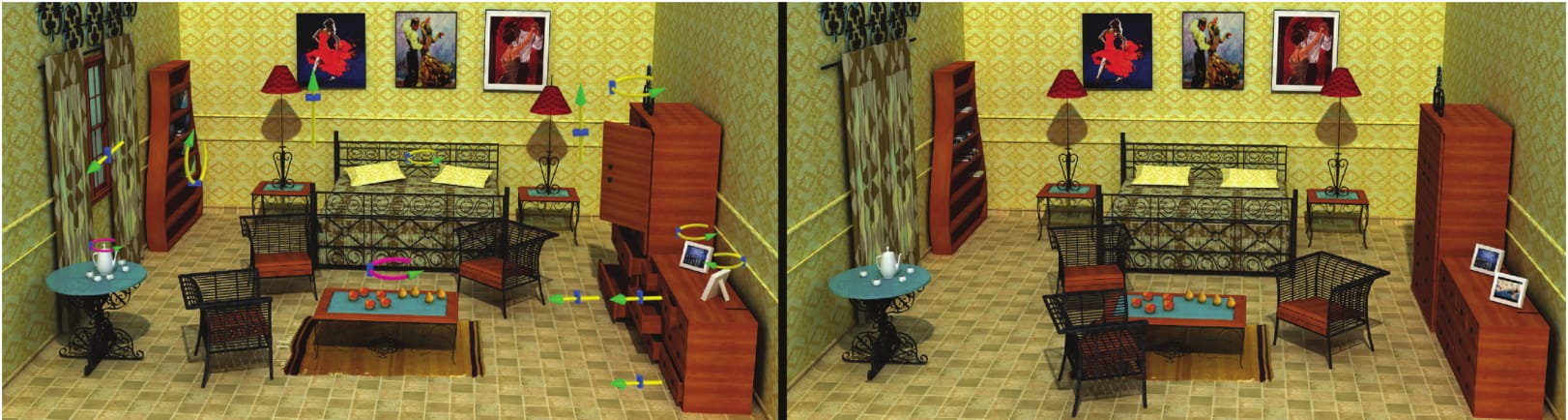}
    \caption{Mobility analysis of 3D scenes~\cite{sharf2014}: mobilities detected in a 3D scene (left) and editing result (right) using the mobility-based controllers (green arrows in the left figure).}
    \label{fig:sharf2014}
\end{figure}

For plausible indoor scenes, the object placements and their spatial relationships are actually derived from \emph{functionalities} of the objects.
For example, chairs are located near tables, lamps are on top of desks, books are organized vertically on shelves.
Functionality analysis of 3D shapes has been studied in~\cite{hu2015, hu2016} for characterizing an object by object-to-object interactions.
In the context of 3D scenes, Sharf et al.~\cite{sharf2014} relate object functionalities to a set of predominant motions that can be performed by the object or its subparts.
These motions are denoted as \emph{mobility}, which defines the specific degrees of freedom, types, axes and limits of motions (Figure~\ref{fig:sharf2014} left).
An object's mobility is computed by analyzing its spatial arrangement, repetitions and relations with other objects and storing it in a \emph{mobility-tree}, which is a high-level functional representation of complex 3D indoor scenes.

Given a static 3D scene, the algorithm segments the scene into object parts and subparts based on their support relations and constructs a support tree hierarchy.
The support relations in the tree infer certain degree of freedom for each supported node with respect to its supporting node (e.g., apples translating on a table).
Repeated instances of the same objects and parts in the scene are detected and their pose differences are analyzed to obtain motion axes and limits.
A set of sophisticated controllers which allow semantical editing operations are defined based on the detected mobilities and used for high-level scene manipulation, such as aligning, reorganizing and deforming mutiple objects or subparts simultaneously (Figure~\ref{fig:sharf2014} right).


\subsection{Analysis of Scene Collections}
\begin{figure}[!htbp] \centering
    \includegraphics[width=0.7\linewidth]{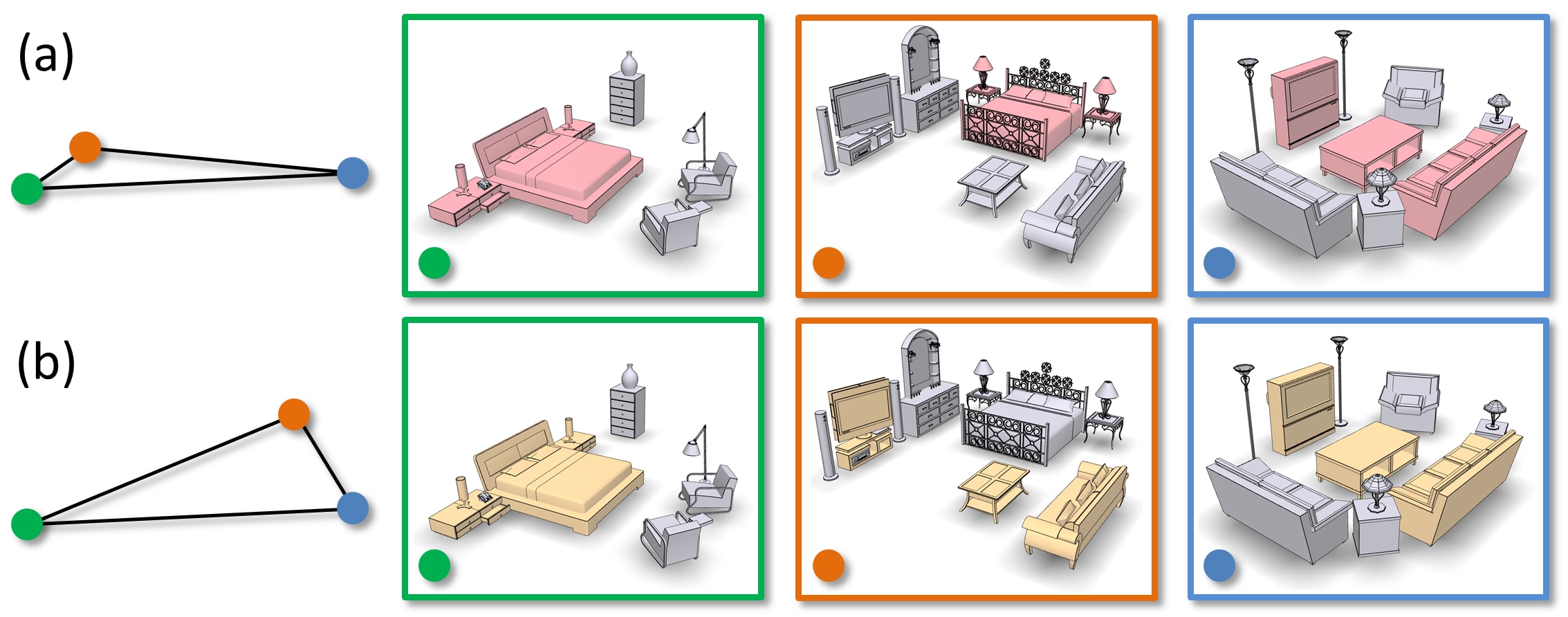}
    \caption{Focal-centric scene similarity: scene comparisons may yield different similarity distances (left) depending on the focal points (sub-scenes in color)~\cite{xu2014}.}
    \label{fig:xu2014_sim}
\end{figure}

Scene collections provide more information about objects and their structural relationships than a single scene.
However, analyzing complex and heterogeneous scenes in a collection is difficult without references to certain points of attention or focus.
Xu et al.~\cite{xu2014} introduce \emph{focal points} for characterizing, comparing, and organizing collections of complex and heterogeneous data and apply the concepts and algorithms developed to collections of 3D indoor scenes.
Focal points are defined as \emph{representative} substructures in a scene collection, by which different similarity distances between scenes could be computed (Figure~\ref{fig:xu2014_sim}).

\begin{figure}[!t] \centering
    \includegraphics[width=\linewidth]{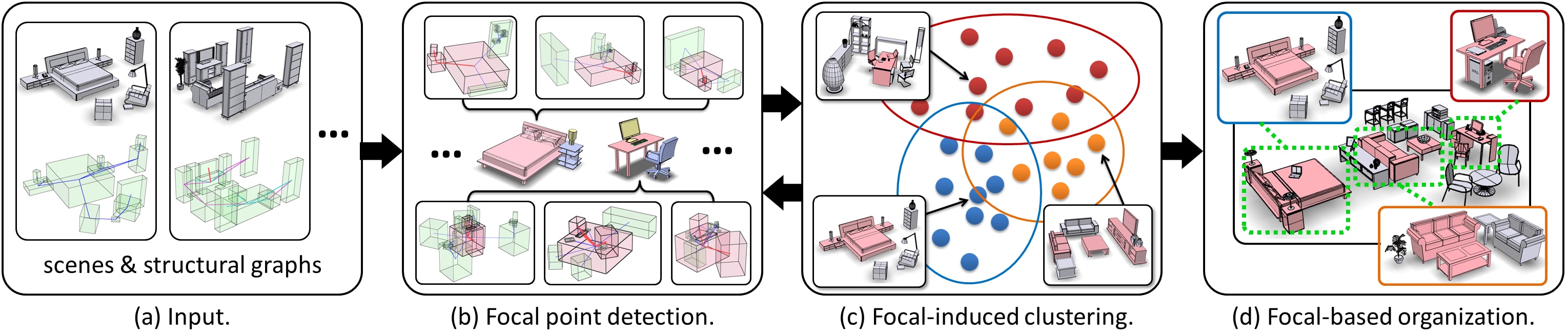}
    \caption{Overview of the focal-driven scene co-analysis~\cite{xu2014}.}
    \label{fig:xu2014}
\end{figure}

To organize a heterogeneous scene collection, the scenes are clustered based on a set of extracted focal points:
scenes in a cluster are closely connected when viewed from the perspective of the representative focal point of that cluster.
For a focal point to be representative, it must occur frequently in the cluster of scenes and also induce a compact cluster.
To solve the coupled problems of focal point extraction and scene clustering,
a \emph{co-analysis} algorithm which interleaves frequent pattern mining and subspace clustering is presented to extract a set of contextual focal points which guide the clustering of the scene collection (Figure~\ref{fig:xu2014}).
In addition to scene organization, applications of the focal-driven scene co-analysis include focal-centric scene comparison and retrieval, and exploration of a heterogeneous scene collection through focal points. 

Graph-based scene representation plays an important role in scene retrieval~\cite{fisher2011} and scene organization~\cite{xu2014}.
Although most of the scenes downloaded from online databases, e.g., 3D Warehouse, are accompanied with scene graphs which are generated during the scene design process, 
the graphs usually lack consistent and semantic object segmentations and category labels.
Pre-processing to consolidate the initial scene graphs are performed in~\cite{fisher2011} so that the graph nodes represent meaningful objects,
but consistent scene hierarchies (e.g., functional groups) are not guaranteed since such information must be inferred from multiple scenes.

\begin{figure}[b!] \centering
    \includegraphics[width=\linewidth]{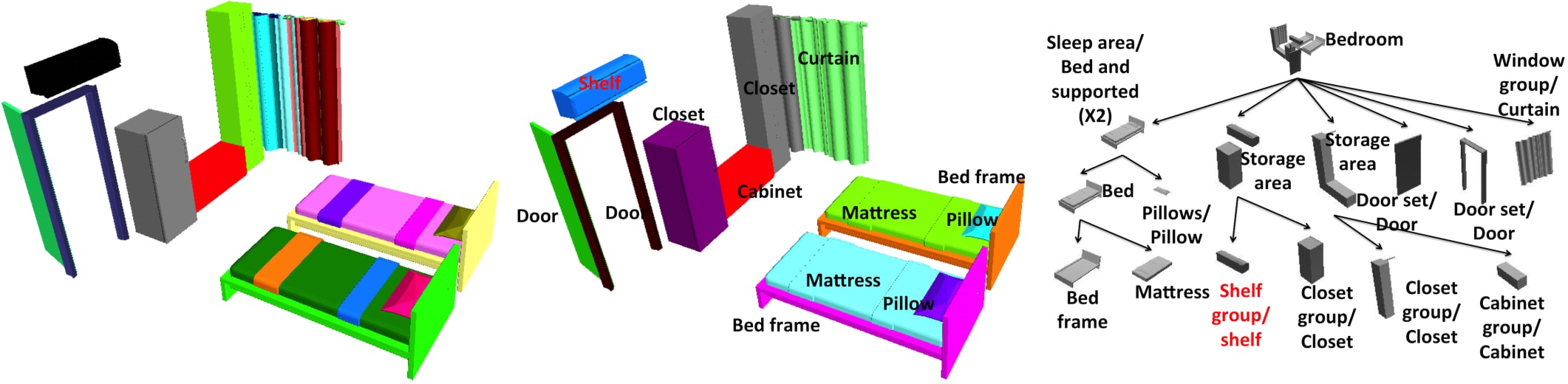}
    \caption{Creating consistent scene graphs using a probabilistic grammar~\cite{xu2014}. An input scene with over-segmentation is parsed into a consistent hierarchy capturing semantic and functional groups (right) as well as semantic object segmentations and labeling (middle).}
    \label{fig:liu2014}
\end{figure}

Liu et al.~\cite{liu2014} develop algorithms that build a consistent representation for the hierarchical decomposition of a scene into semantic components (Figure~\ref{fig:liu2014}).
Given a collection of consistently-annotated scene graphs representing a category of scenes (e.g., bedroom, library, classroom, etc.) as the training set, a probabilistic hierarchical grammar that captures the scene structure is learned.
Next, the learned grammar is used to hierarchically segment and label novel scenes and create consistent scene graphs.
The resultant representation not only segments scenes into objects with semantic classes, but also establishes functional roles and relationship of objects (e.g., \emph{dining} table, \emph{study} table, table-\emph{and}-chairs) which could be used for functional scene understanding.

	\section{Modeling and Synthesis of 3D Indoor Environments}

With widely applications in 3D games, interior design and VR/AR, there is an increasing demand for 3D indoor scenes, especially those with complex object composition and arrangements.
However, creating 3D scenes is difficult for novice users and designing of a detailed and realistic scene is time-consuming even for a professional artist.
To solve the bottleneck of content-creation for indoor scenes, many efforts have been made in the graphics and vision communities to develop intelligent tools and methods for 3D scene modeling and synthesis.

To generate a scene, two questions usually need to be answered: what objects and how many of them should appear in the scene;
where an object should be placed in the scene.
Such knowledge could be extracted from various existing scene data~\cite{fisher2012, silberman2012, lin2014, song2015},
thus enables data-driven 3D scene generation.

When learning and modeling objects and their relationships, one key concept is the \emph{context}, which depicts the surrounding environment of an object in the scene.
The context provides rich information about how an object is interacting with its neighbors and could be used for inferring the occurrence and arrangement of the object.
Context-based modeling is explored for interactive scene modeling~\cite{yu2016, savva2017} as well as automatic scene synthesis~\cite{fisher2012, zhao2016}.
More recently, human-object context which has been studied in computer vision and robotics for human-centric scene understanding is introduced to graphics for human-centric 3D scene modeling. 

In this chapter, we first review suggestive tools developed for interactive scene modeling~\cite{fisher2010, yu2016, savva2017}.
Next, we study different modeling techniques for creating 3D scenes based on multitude inspirations such as text~\cite{chang2014b, chang2017}, images~\cite{liu2015, izadinia2017} and 3D scene exemplars~\cite{fisher2012}.
Finally, we introduce human-centric 3D scene modeling which exploits the human-object context for scene analysis~\cite{savva2014} and synthesis~\cite{fisher2015, ma2016}.

\subsection{Suggestive Scene Modeling}
Interactive and user-centric tools for 3D scene modeling are available online, e.g., \emph{Autodesk Homestyler}~\cite{homestyler}, \emph{Sweet Home 3D}~\cite{swhome}, \emph{Planner 5D}~\cite{planner5d}.
The use of such tools requires advanced modeling skills and the modeling time is often quite long.
In research community, early work by Xu et al.~\cite{xu2002} used a set of intuitive placements constraints such as non-interpenetration of objects, object stability to allow the simultaneous manipulation of a large number of objects.
Recent works have explored the interactive context-based model suggestion~\cite{fisher2010, yu2016, savva2017}, by which scene modeling could be simplified as a set of point-and-click operations~\cite{savva2017}:
a user just needs to specify a region in a scene and select an object from a list of candidates which are retrieved based on the current context of the scene; the object is then placed and oriented in the scene automatically.

\begin{figure}[!t] \centering
    \includegraphics[width=\linewidth]{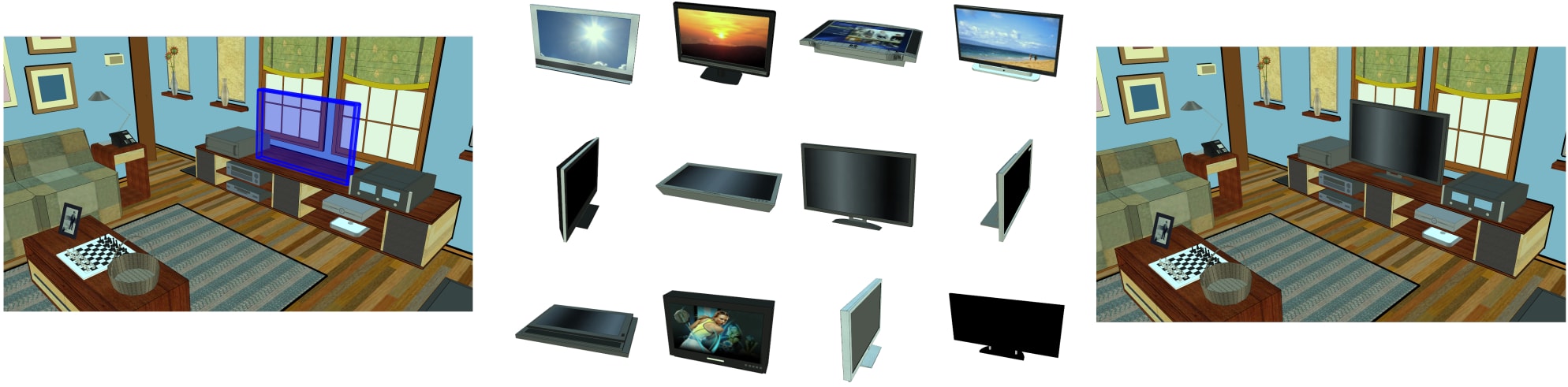}
    \caption{Scene modeling using a context-based search~\cite{fisher2010}: a user places the blue box in the scene and asks for models that belong at this location (left); the system returns models from database that match the provided neighborhood (middle); the user selects a model from the list and it is inserted into the scene (right).}
    \label{fig:fisher2010}
\end{figure}
Traditional ways of adding an object into a scene include first obtaining a 3D model of the object using keyword-based search or selection from a hierarchical list which organizes objects into semantic categories, and then inserting the model into the scene.
In this process, the tasks of 3D model search and scene modeling are actually decoupled.
The work of Fisher and Hanrahan~\cite{fisher2010} develops a context-based 3D model search engine, which changed the order of scene modeling operations from first searching and then placing, to first locating and then searching (Figure~\ref{fig:fisher2010}).
The context search uses a data-driven approach, learning object co-occurrences and their spatial relationships from existing scenes.
Given the query box where the user wants to insert a new object, the algorithm ranks each object in the database according to how well it fits to the box as well as the its relationships to the context objects.
In Fisher et al.~\cite{fisher2011}, context-based model search is also achieved using graph kernel and the performance is improved since more semantic and structural relationships between objects are considered. 

\begin{figure}[!t] \centering
    \includegraphics[width=0.9\linewidth]{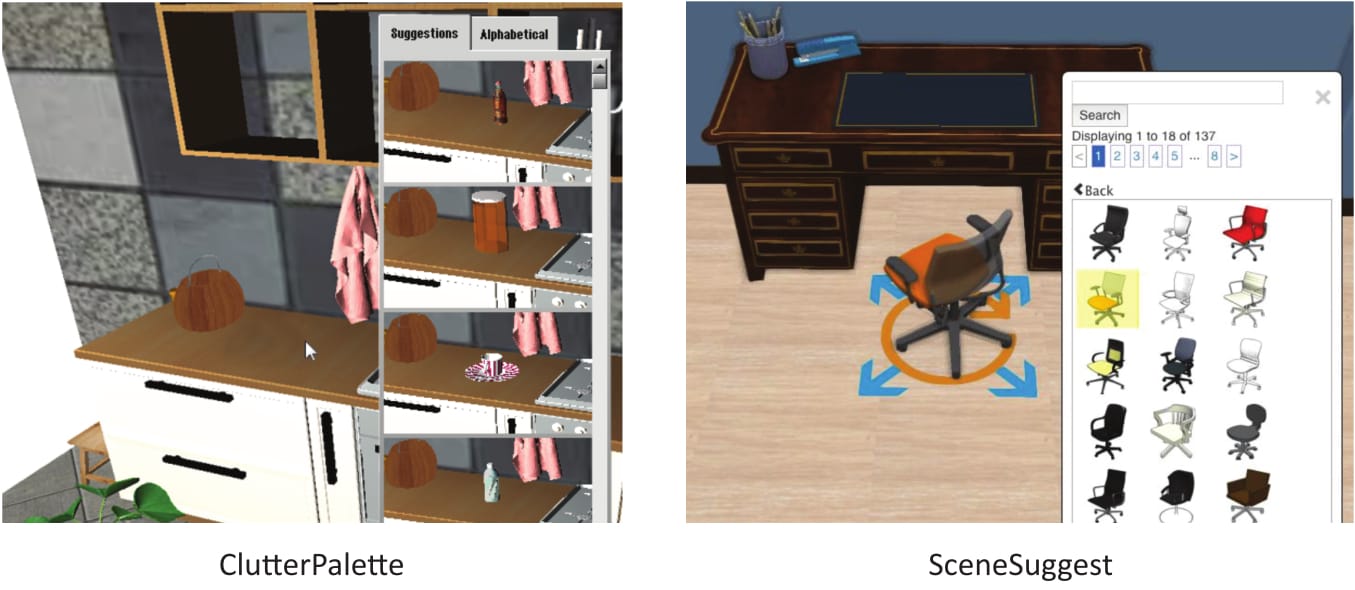}
    \caption{Suggestive scene modeling interface of ClutterPalette~\cite{yu2016} and SceneSuggest~\cite{savva2017}.}
    \label{fig:interface_yu_savva}
\end{figure}
Instead of focusing on model retrieval, interactive scene modeling systems have been proposed in~\cite{yu2016, savva2017} (Figure~\ref{fig:interface_yu_savva}).
Yu et al.~\cite{yu2016} introduce the ClutterPalette, an interactive tool for detailing indoor scenes with small-scale items which are critical but often missing in existing 3D scenes.
When a user points to a location in the scene, the ClutterPalette suggests detail items for that location.
In order to present appropriate suggestions, the Clutterpalette is trained on NYU Depth Dataset~\cite{silberman2012}, a dataset of RGB-D images of real-world scenes, annotated with semantic objects and their support relations.
Co-occurrences of objects are extracted from the dataset and used to provide suggestions based on objects that have already been added into the scene.
The recent system SceneSuggest presented by Savva et al.~\cite{savva2017} improves ClutterPalette by learning continuous distributions of object position, orientation from existing 3D scenes~\cite{fisher2012} and combining the priors with the context of a user specified region to suggest a list of relevant 3D models.
Using more accurate relative distances and orientations extracted from synthetic 3D data, SceneSuggest is able to place and orient objects automatically rather than manual adjustment in~\cite{fisher2010, yu2016}.

\subsection{Scene Modeling from X}
Data-driven scene modeling or reconstruction from X has gained much interest lately where X could be a sketch~\cite{xu2013}, a photograph~\cite{liu2015,izadinia2017}, or some text descriptions~\cite{chang2014b, chang2017}.
Semantic modeling from captured 3D data reviewed in Chapter~\ref{chap:scan} can also be classified into this category.
In these cases, X provides inspirations and a target for scene modeling.
The objects and their arrangements are inferred from X to guide the retrieval and placement of suitable 3D objects from a model repository.

\begin{figure}[!t] \centering
    \includegraphics[width=\linewidth]{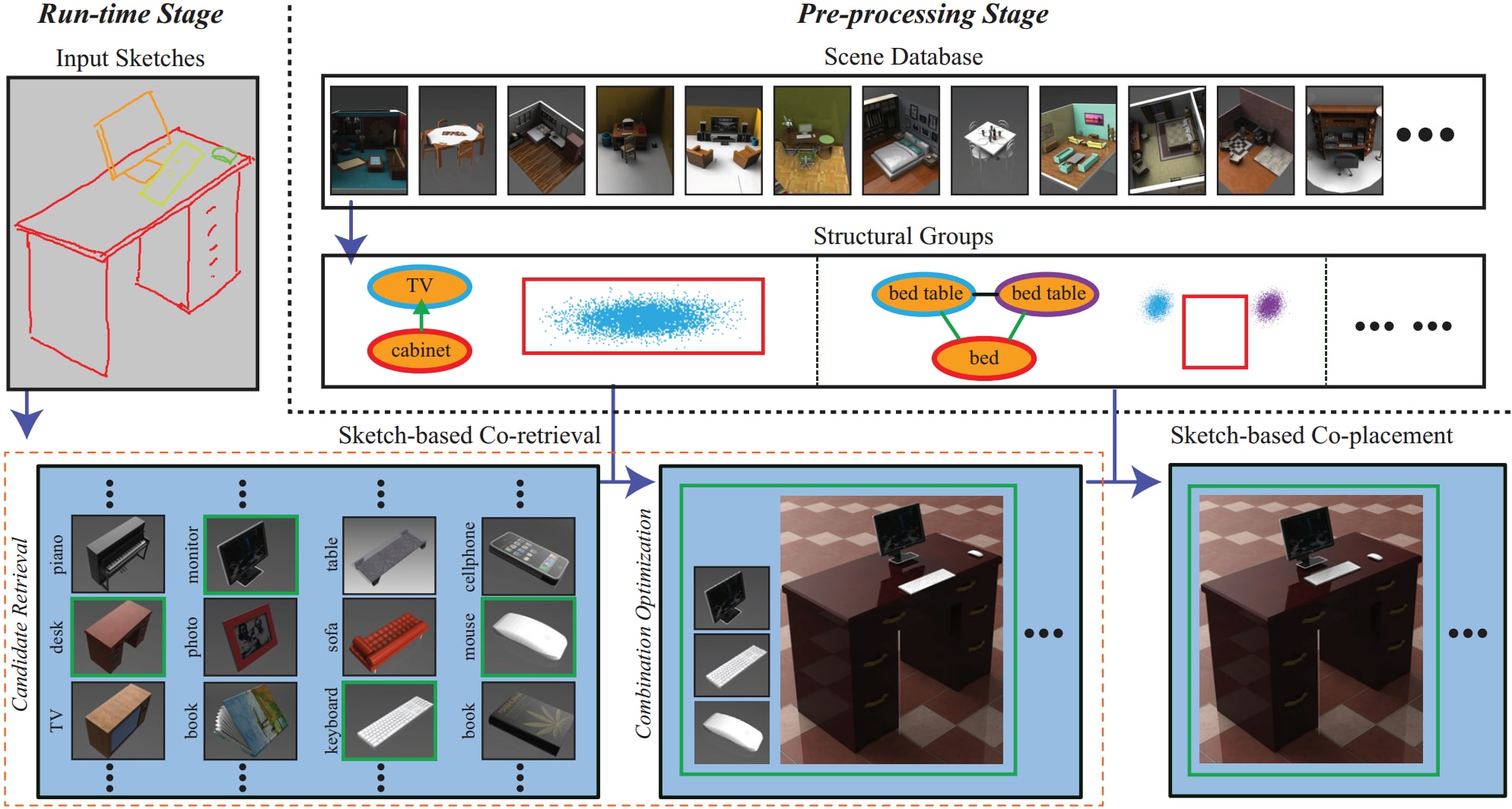}
    \caption{Pipeline of Sketch2Scene~\cite{xu2013}. The system pre-processes a large database of well-constructed 3D scenes to extract a small set of structural groups, which are used at runtime to automatically convert input segmented sketches to a desired scene.}
    \label{fig:xu2013}
\end{figure}

The availability of large collections of 3D models (e.g., 3D Warehouse) together with various shape retrieval techniques offers a great opportunity for easy composition of new 3D scenes or models by properly recombining the existing models or their
parts.
Sketch-based user interface is commonly adopted for this task mainly due to its simplicity, intuitiveness, and ease of use~\cite{eitz2012}.
Early works about sketch-based scene modeling~\cite{shin2007,lee2008} typically repeat the following process for \emph{individual} objects one by one: first input a 2D sketch of an object, then retrieve and place a 3D model that best matches the input sketch.
Focusing on joint processing of a set of sketched objects, Xu et al.~\cite{xu2013} present Sketch2Scene, a framework that automatically turns a sketch inferring multiple scene objects to semantically valid, well arranged scenes of 3D models (Figure~\ref{fig:xu2013}).
Taking a set of segmented sketches as input, the system~\cite{xu2013} performs \emph{co-retrieval} and \emph{co-placement} of relevant 3D models enabled by \emph{structural groups}, a compact summarization of reliable relationships among objects in a large database of well-constructed 3D scenes.
Constraints from the input sketches as well as the structural groups are combined together to produce 3D scenes with no user intervention.

Comparing to sketches, 2D images and photographs of indoor scenes contain more rich object compositions which could inspire the 3D scene modeling or reconstruction.
There are several problems for 3D indoor scene modeling from a single image:
object detection and segmentation from cluttered indoor environments;
3D model retrieval and alignment based on an incomplete, single-view image region;
3D recovery of the room geometry and structure.
With the advancement of computer vision algorithms for these problems, 2D image to scene generation is tackled in~\cite{liu2015, izadinia2017} recently.
Given a single 2D image depicting an indoor scene as input, Liu et al.~\cite{liu2015} reconstruct a 3D scene in two stages:
image analysis which obtains object segmentations and their 3D location using~\cite{lee2009};
3D model retrieval based on matching the line drawings extracted from 2D objects and 3D models.
As the object segmentation and 3D recovery in~\cite{liu2015} are based on a set of limited assumptions, e.g., easy-to-segment objects, regular room structure, the system can only reconstruct simple 3D scenes (sofa, chair, coffee table etc.) from clean indoor images.

\begin{figure}[!htbp] \centering
    \includegraphics[width=\linewidth]{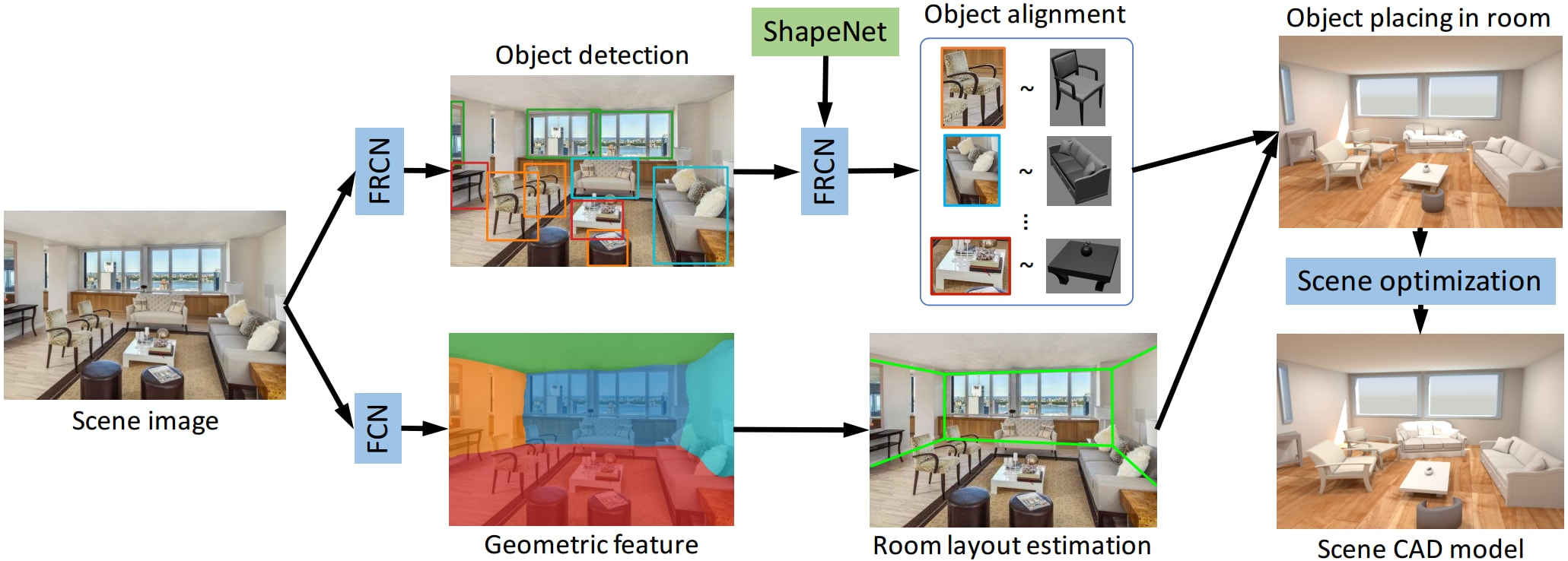}
    \caption{Overview of IM2CAD~\cite{izadinia2017}: an input image (left) is processed through a series of steps to produce a scene CAD model (bottom right).}
    \label{fig:izadinia2017}
\end{figure}
Using the state-of-the-art object recognition and scene understanding algorithms, Izadinia et al.~\cite{izadinia2017} present IM2CAD which automatically produces high-quality scene modeling results on challenging photos from interior home design websites(Figure~\ref{fig:izadinia2017}).   
Instead of using handcrafted features as in~\cite{liu2015}, the IM2CAD leverages deep features trained by convolutional neural nets (CNNs)~\cite{krizhevsky2012} to reliably match between photographs and CAD renderings extracted from ShapeNet~\cite{chang2015a}.
The system iteratively optimizes the placement and scales of objects to best align scene renderings to the input photo, operating jointly on the full scene at once, and accounting for inter-object occlusions.
Although IM2CAD obtains significant improvements in the 3D room layout estimation as well as 3D object location from single RGB images, the resulting scenes are still too simple that only large furnitures, walls and windows could be modeled.

Natural language is arguably the most accessible input for content creation.
Generating 3D scenes from text has been a long and on-going pursuit since the pioneering work on WordsEye~\cite{coyne2001}.
WordsEye, along with other follow-ups~\cite{seversky2006, coyne2012} is capable of generating 3D scenes directly from natural language descriptions, but relies on manual mapping between explicit scene arrangement languages and object placements in a scene, e.g., ``the chair is three feet to the north of the window''.
A series of papers by Chang et al.~\cite{chang2014a, chang2014b, chang2017} have provided improvements over the early text-to-scene systems by utilizing spatial knowledge, learned from 3D scene data, to better constrain scene generations with unstated facts or common sense.

\begin{figure}[b!] \centering
    \includegraphics[width=\linewidth]{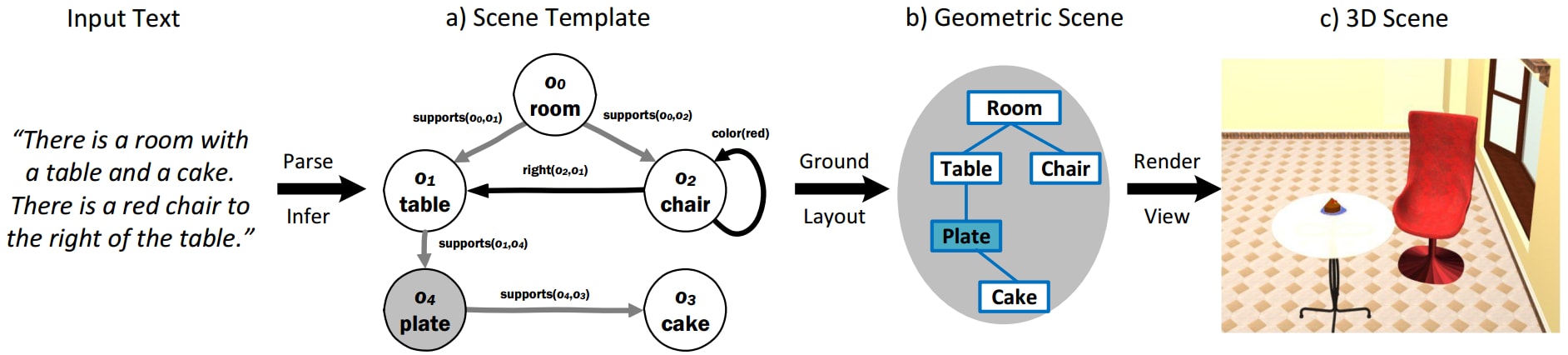}
    \caption{Text-to-scene generation pipeline in Chang et al.~\cite{chang2014b}: the input text is parsed into a scene template with implicit spatial constraints inferred from learned priors; the template is then grounded to a geometric scene, by instantiating and arranging 3D models into a final 3D scene.}
    \label{fig:chang2014b}
\end{figure}
In the representative work of Chang et al.~\cite{chang2014b}, the input text is first parsed into a \emph{scene template}, a graph representation that captures semantics and spatial relations of a scene.
To infer spatial knowledge that is often not expressed in natural language, priors including object occurrence, static support relations and relative positions of objects are learned from the 3D scene database~\cite{fisher2012}
and used for enriching the scene template with implicit constraints.
The scene template is then grounded into a geometric 3D scene by querying a 3D model database and arranging the objects based on the constraints encoded in the template and the spatial knowledge priors.
Their latest SceneSeer system~\cite{chang2017} further extends the pineline in~\cite{chang2014b} with interactive text-based scene editing operations, e.g., adding, removing, replacing objects, and mouse-based scene manipulation to refine a generated scene.

\subsection{Furniture Layout Optimization}
One line of work for 3D scene generation focuses on furniture layout optimization~\cite{germer2009, merrell2011, yu2011} for a given room with a given set of furniture or objects.
The common problem is to automatically generate object arrangements that satisfy a set of indoor scene layout constraints or rules.

Germer and Schwarz~\cite{germer2009} arrange a room by letting each piece of furniture act as an agent in a multi-agent system, following manually specified room semantics and furniture layout rules. 
By using the scheme of procedural furniture arrangements, the system of~\cite{germer2009} is suitable for creating room interiors of complex and interactive environments such as furnishing rooms of a building while real-time walkthroughs.

\begin{figure}[!htbp] \centering
    \includegraphics[width=0.7\linewidth]{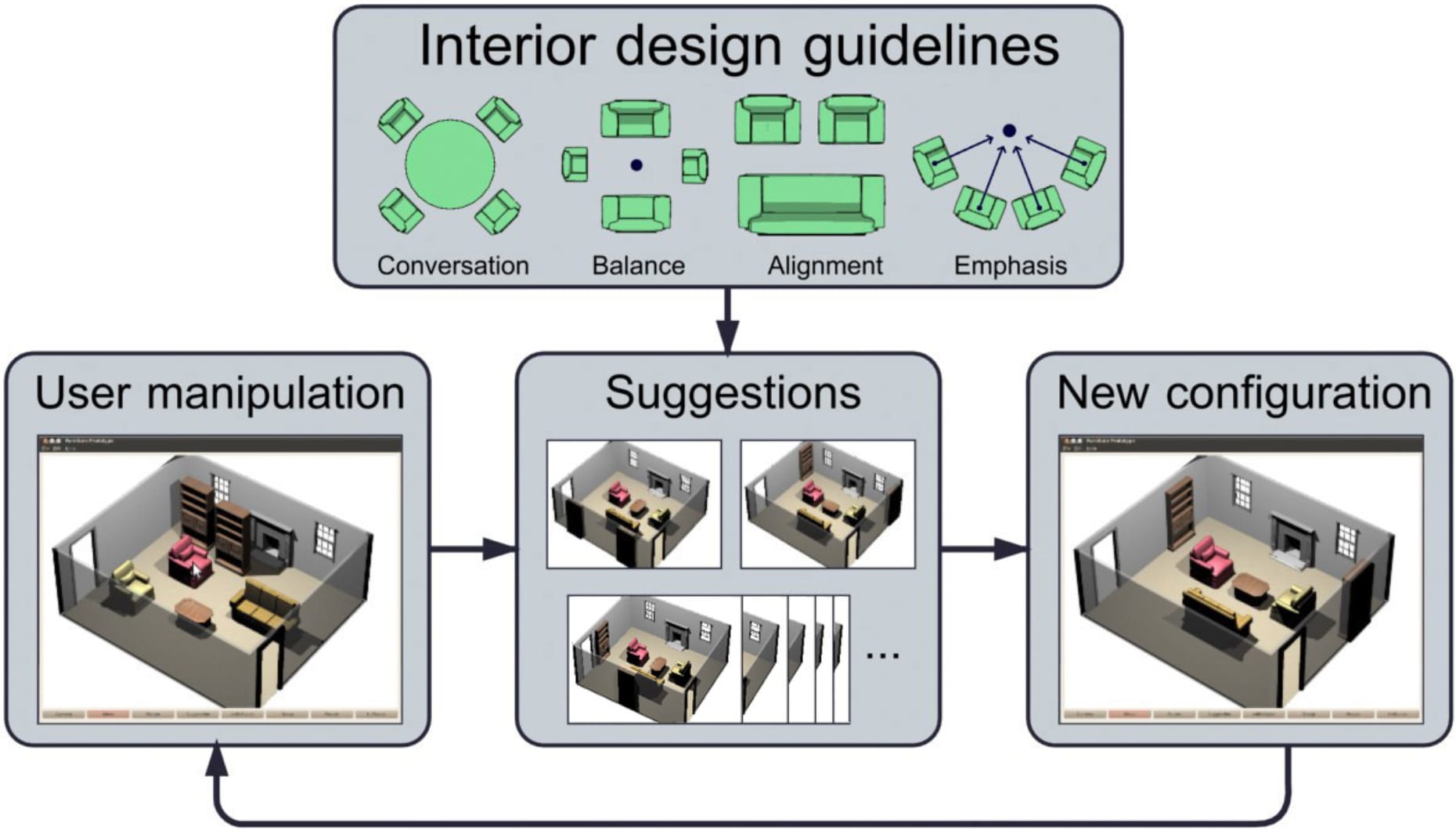}
    \caption{Interactive furniture layout system in~\cite{merrell2011}. In response to user manipulation, the system suggests new arrangements that respect user-specified constraints and follow interior design guidelines.}
    \label{fig:merrell2011}
\end{figure}

Merrell et al.~\cite{merrell2011} present an interactive furniture layout system that assists users
by suggesting furniture arrangements that are based on interior design guidelines (Figure~\ref{fig:merrell2011}).
The set of furniture layout guidelines such as circulation, balance and alignment, are identified from interior design manuals and consultation with professional designers.
These guidelines are encoded as terms in a density function and layout suggestions are generated by sampling this function while respecting user's constraints.

Similar to~\cite{merrell2011}, the \emph{Make it Home} system of Yu et al.~\cite{yu2011} encodes spatial relationships for furniture objects into a cost function and automatically synthesizes furniture layouts by optimizing the function.
The difference is the spatial relationships of objects, such as relative distance and orientation as well as support relations are learned from 3D scene exemplars instead of manual specification in~\cite{merrell2011}. 
Also, to make sure the generated furniture arrangements are functional and realistic, factors including accessibility, visibility and pathway connecting the doors are added to the cost function.
As the search space for optimizing the function is highly complex, simulated annealing is applied to search for a good approximation to the global optimum, and realistic furniture arrangements are produced as the results.

Comparing to generating layouts with fixed objects instances, the automatic \emph{open world} layout synthesis is more appealing as the goal is to generate diverse layouts with unspecified number of objects.
Yeh et al.~\cite{yeh2012} use factor graphs, a type of graphical model, to encode complex object relationships as constraints, and propose a variant of Markove Chain Monte Carlo (MCMC) method to sample the possible layouts.
The algorithm succeeds at synthesizing interior environments, such as coffee shops, with varying shapes and
scales.
However, the resultant scenes are restricted with patterns that are only handedcrafted in the code.
 
\subsection{3D Scene Synthesis}

\begin{figure}[!htbp] \centering
    \includegraphics[width=0.8\linewidth]{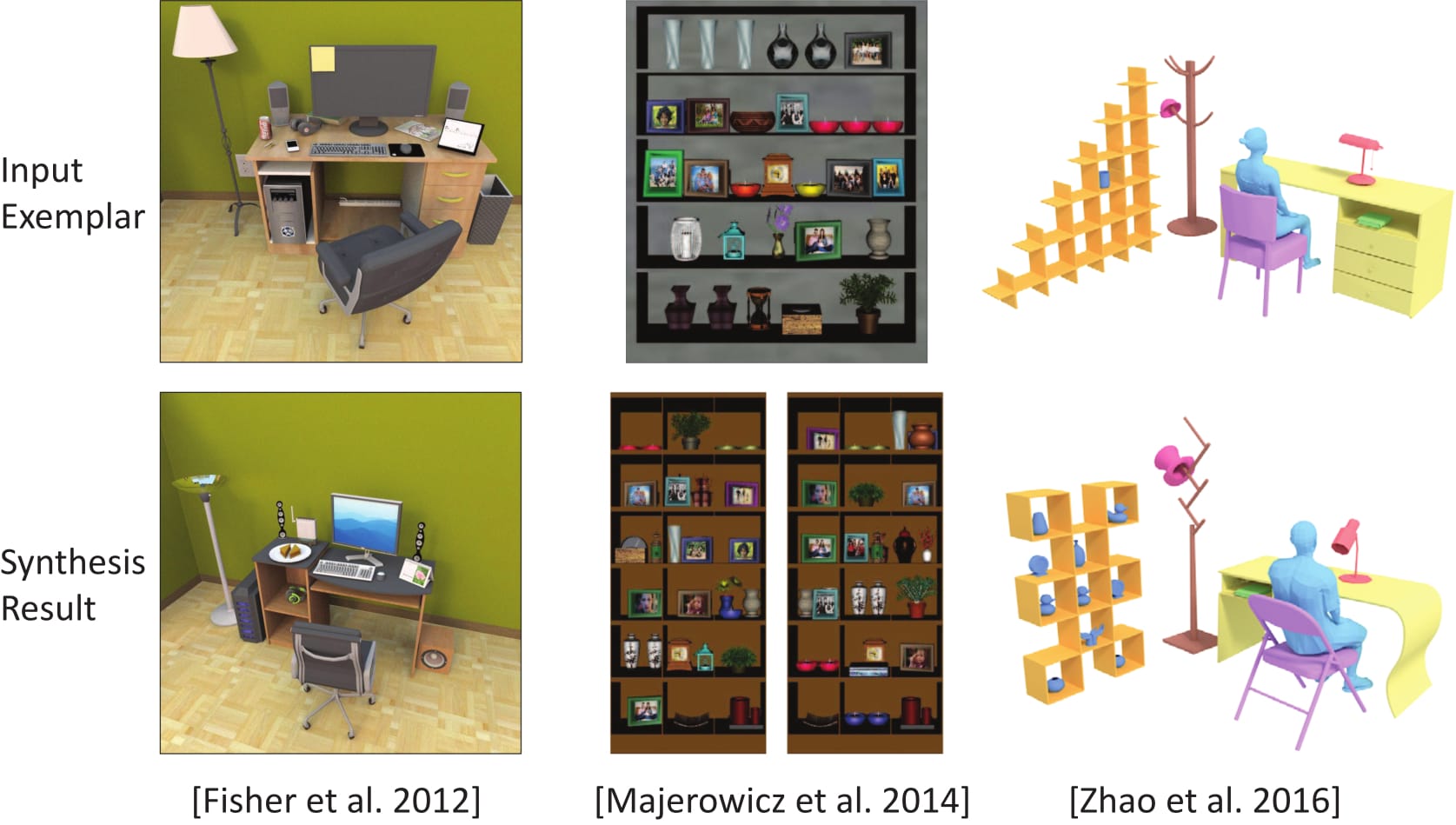}
    \caption{Different example-based scene synthesis approaches. Scenes with similar object occurrences and relationships (left), arrangement style (middle), and complex relationships (right) are synthesized based on the input exemplars.}
    \label{fig:example_based}
\end{figure}

In contrast to 3D scene modeling from observations, 3D scene synthesis aims to generate \emph{novel} scenes with plausible object compositions and arrangements.
Given only one or a few exemplars which could be 3D scenes or images, example-based approaches~\cite{fisher2012, majerowicz2014, zhao2016} perform probabilistic reasoning from exemplars and 3D scene databases to synthesize scenes that are similar to the input, but with certain diversity (Figure~\ref{fig:example_based}).
Recently, progressive scene synthesis~\cite{zeinab2016} is proposed to synthesize 3D scenes by inserting one or more objects progressively into the scene.

The best known method for synthesizing realistic 3D scenes is the example-based scene synthesis proposed by Fisher et al.~\cite{fisher2012}.
Given a few user-provided examples, the system can synthesize a diverse set of plausible new scenes by learning from a
3D scene database.
The main challenge for example-based synthesis is generating a \emph{variety} of results while retaining plausibility and similarity to the examples.
Two ways are exploited in~\cite{fisher2012} to improve the diversity of synthesized scenes:
firstly, extract \emph{contextual categories} from the scene database by using object neighborhoods in scenes to group together objects that are likely to be considered interchangeable;
secondly, treat the scene database as a prior over scenes and train a probabilistic model on both the examples and relevant scenes retrieved from the database using~\cite{fisher2011}.
The probabilistic model consists of an occurrence model, which specifies what objects should appear in the generated scenes, and an arrangement model, which specifies where those objects should be placed.
Using the contextual categories and the \emph{mixed} probabilistic model, the algorithm successfully synthesizes scenes with a large variety of both objects and arrangements.

Focusing on enriching scenes with more details, i.e. small objects on the shelves,
Majerowicz et al.~\cite{majerowicz2014} present an example-based method which automatically populates empty shelf-like surfaces with diverse arrangements of artifacts in a given style.
Here, the style of an arrangement is defined as a combination of object-level measures, which capture local arrangement properties, such as the percentage of instances of a particular object and the relative location of objects,
and global measures, which compare high-level characteristics of the two arrangements, such as density and symmetry.
Given an annotated photograph or a 3D model of an exemplar arrangement that reflects the desired context and style,
the algorithm learns the style from this exemplar, and generate diverse arrangements by stochastic optimization.

Real world indoor scenes contain complex inter-object relations, e.g., one object is ``hooked on'', ``surrounded by'' or ``tucked into'' another object.
To generate scenes with such complex relations, Zhao et al.~\cite{zhao2016} take an example-based approach to synthesize new scenes by replacing objects in the example scene with database objects while preserving the original complex spatial relations in the example.
The scene synthesis problem is modeled as a template fitting process, where a \emph{relationship template} encoding rich spatial relationships between objects is computed from an example scene and new scene variations are synthesized by fitting novel objects to the template.
The relationship template is defined by combining the IBS~\cite{zhao2014} with a novel feature called the Space Coverage Feature (SCF), which describes the spatial relations between points in open space and surrounding objects.
Given a relationship template extracted from the example scene, novel objects are fitted to the template by matching the nearby open space to the features stored in the template to generate new scenes.


Unlike the example-based methods which take a holistic view of scene generation by targeting overall similarity between the generated scene and the exemplars,
Sadeghipour et al.~\cite{zeinab2016} propose the progressive scene synthesis which inserts objects progressively into an initial scene based on probabilistic models learned from a large-scale annotated RGB-D dataset, in particular, the SUN RGB-D databset~\cite{song2015}, which consists of more than 10,000 depth images of real scenes.
Similar to~\cite{fisher2012}, the probabilistic models are also summarized as a \emph{co-occurrence} model and an \emph{arrangement} model.
In addition to considering pairwise object relations, higher-oder relations including symmetry, distinct orientations and proximity are learned from the database.
The objects along with their relations extracted from the database scenes are encoded in a factor graph from which the new objects are sampled based on existing objects.
By progressively selecting and placing objects, the scene synthesis process attains more local controllability and ensures global coherence and holistic plausibility of the synthesized scenes.

\subsection{Human-centric Scene Modeling}
We live in a 3D world, performing activities and interacting with objects in the indoor environments everyday.
It is easy for humans to understand the surrounding scenes or arrange objects based on their functionality.
However, it is challenging for an agent, e.g., a robot, to automatically generate behaviors and interact with the 3D environments since the robot lacks knowledge of the objects as well as their functionalities. 
There has been a great deal of work in robotics and computer vision on utilizing human-centric approaches in the scene analysis and modeling tasks, e.g., scene geometry estimation~\cite{fouhey2012}, object labeling ~\cite{jiang2012b}, robot placement~\cite{jiang2012b}, just to name a few.
Recently, human-object context are exploited in computer graphics for functional scene understanding~\cite{savva2014}.
For scene generation, human actions or activities are used to guide the scene synthesis from depth scans~\cite{fisher2015} or drive the evolution of object arrangements in the scene~\cite{ma2016}.
By using a human-centric perspective, indoor scene modeling focuses more on the functionality and plausibility of the results, making one step closer to the generation of realistic scenes.

\begin{figure}[!htbp] \centering
    \includegraphics[width=\linewidth]{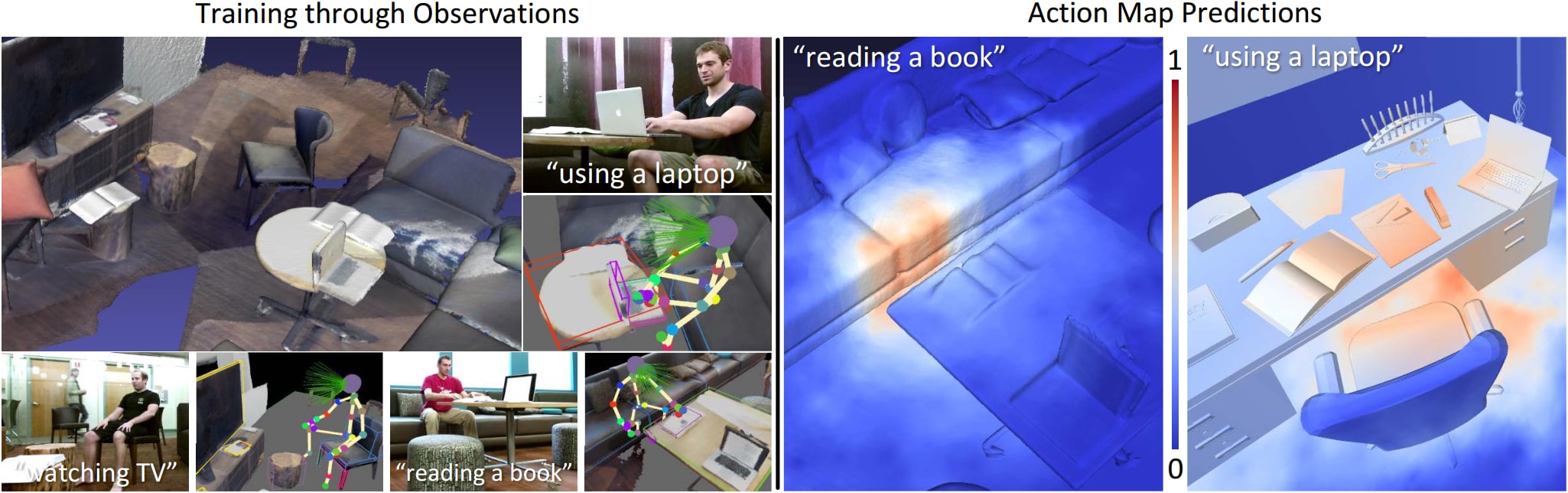}
    \caption{Overview of SceneGrok~\cite{savva2014}: the system starts by scanning the geometry of real environments using RGB-D sensors (left); with the observations of people interacting with the captured environments (mid-left), a classifier which transfers interaction knowledge to unobserved scenes is trained and used for predicting action maps, the likelihood of actions occurring in regions, in scanned or virtual scenes (mid-right and right, respectively).}
    \label{fig:savva2014}
\end{figure}

While a large amount of 3D scenes exist, the semantics and functionality of the objects and scene regions are usually missing.
The work of Savva et al.~\cite{savva2014} present a method to establish a correlation between the geometry and the functionality of 3D environments.
The idea is to learn the functionality or affordance of 3D scenes directly from observations of people interacting with real environments captured by RGB-D sensors (Figure~\ref{fig:savva2014}). 
By observing humans performing different actions, e.g., ``use a desktop PC'', correlations between human body poses and the surrounding scene geometry are extracted to train action classifiers which can transfer interaction knowledge to unobserved scenes.
Given a new, unobserved scene, action maps which encode the probability of specific actions taking place
in the scene are predicted based on the classifiers and the scene geometry.

\begin{figure}[!htbp] \centering
    \includegraphics[width=\linewidth]{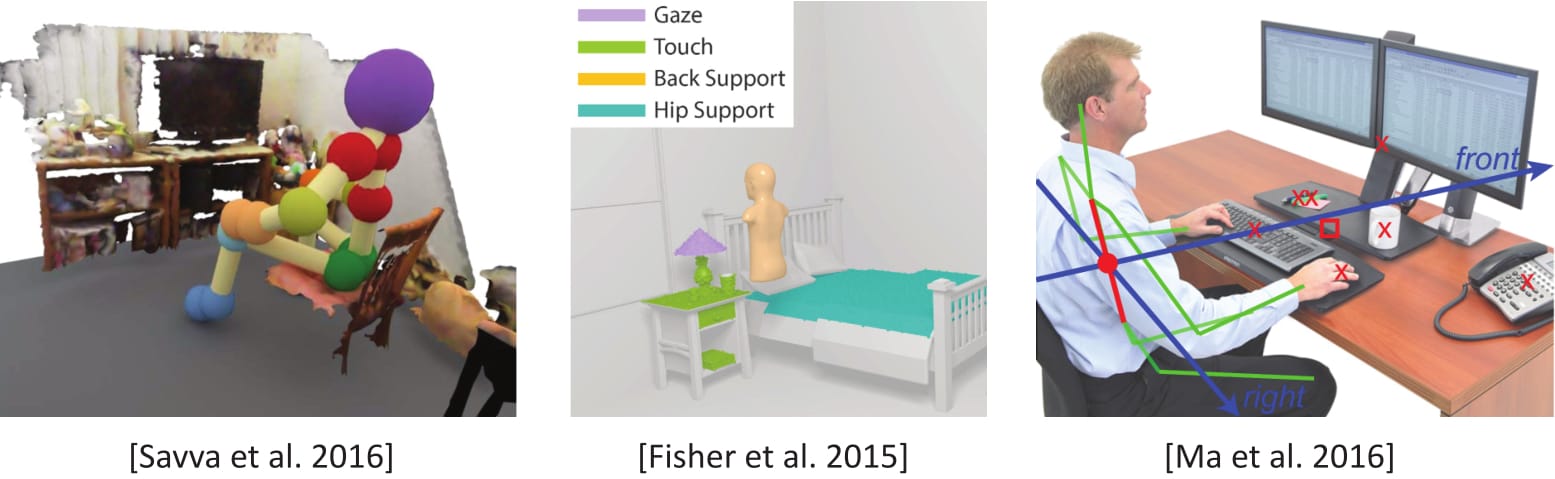}
    \caption{Different types of annotated training data for learning action or activity models. Object distributions from RGB-D scans (left), 3D scenes (middle) and photos (right) are extracted to learn the probabilistic models of corresponding actions, ``sit on a chair and watch tv'', ``prepare for bed'', ``use computer on desk while sitting'' (from left to right).}
    \label{fig:action_models}
\end{figure}

Instead of learning the supervised action classifiers, recent works build probabilistic models for human-object interaction, i.e., actions or activities, and learn the parameters from multitude data sources, such as depth scans, 3D scenes and photographs (Figure~\ref{fig:action_models}).
The learned models are further utilized for generating human-object interaction snapshots~\cite{savva2016} or scene synthesis~\cite{fisher2015, ma2016}.

\begin{figure}[!htbp] \centering
    \includegraphics[width=\linewidth]{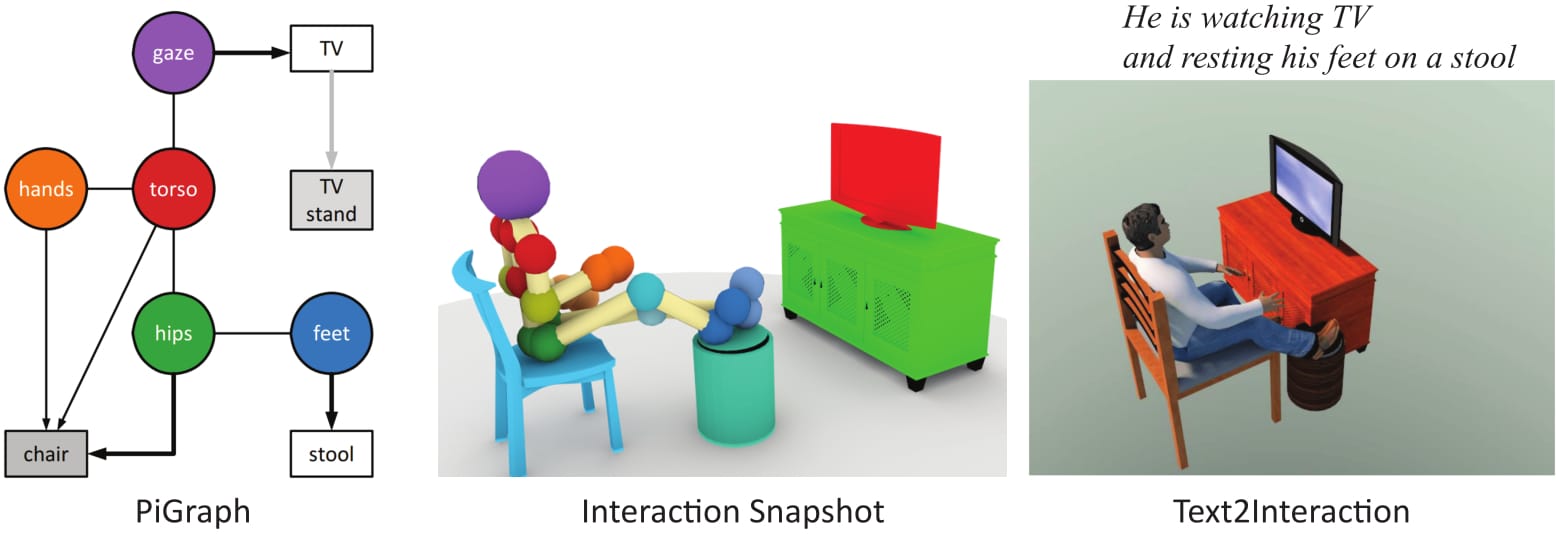}
    \caption{The PiGraph representation (left) and its applications for generating interaction snapshot (middle) and text2interaction (right) in~\cite{savva2016}.}
    \label{fig:pigraph}
\end{figure}

Following their work for action-based scene understanding, Savva et al.~\cite{savva2016} learn a probabilistic model connecting human poses and arrangements of object from real-world observations captured using the same framework as in~\cite{savva2014}.
The model is represented as a set of \emph{prototypical interaction graphs} (PiGraphs) (Figure~\ref{fig:pigraph} left), a human-centric representation capturing physical contact and visual attention linkages between 3D geometry and human body parts.
The joint probability distributions over pose and object geometry encoded in the PiGraphs are used to generate \emph{interaction snapshots} (Figure~\ref{fig:pigraph} middle), which are static depictions of human poses and relevant objects during human-object interactions.
The priors learned for PiGraphs also have applications such as text2interaction (Figure~\ref{fig:pigraph} right).

\begin{figure}[!htbp] \centering
    \includegraphics[width=\linewidth]{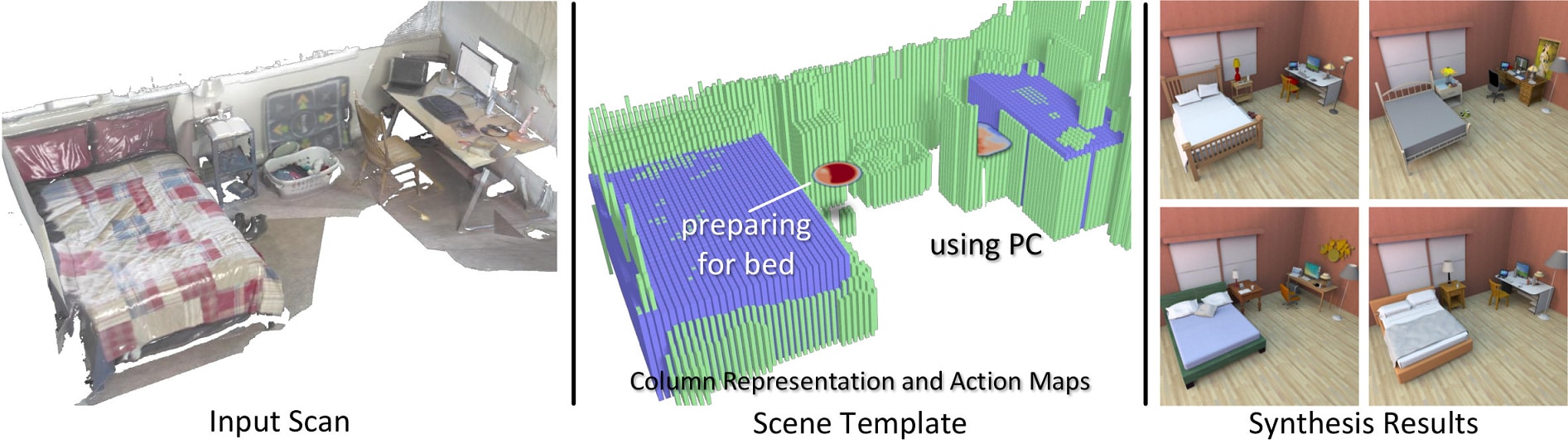}
    \caption{Activity-centric scene synthesis~\cite{fisher2015}. Given a 3D scan of an environment (left), a scene template which encodes where activities can occur and contains a coarse geometric representation describing the general layout of objects is produced (middle); plausible scenes (right) are synthesized based on the scene template.}
    \label{fig:fisher2015}
\end{figure}
Targeting at the generation of realistic 3D scenes, Fisher et al.~\cite{fisher2015} present a novel method to synthesize scenes that allow the same activities as real environments captured through noisy and incomplete 3D scans (Figure~\ref{fig:fisher2015}).
As robust object detection and instance retrieval from low-quality depth data is challenging~\cite{nan2012, kim2012, shao2012, chen2014}, the algorithm aims to model semantically-correct rather than geometrically-accurate object arrangements.
Based on the key insight that many real-world environments are structured to facilitate specific human activities, such
as sleeping or eating, the system anchors an observed scene in a set of activities and uses associated activity regions as focal points to guide scene modeling.

Given a 3D scan of an environment as input, a \emph{scene template} which captures both the prominent geometric properties of the scan and the activities that are likely to be performed in the environment is computed.
In order to synthesize object arrangements based on the activities predicted in the template,
activity models which encode distributions of objects involved in specific activities are trained from an annotated 3D scene and model database in a pre-processing step.
In the synthesis step, objects are iteratively placed in the scene based on the the geometric and activity properties specified in the template, while the activity properties are measured by the learned activity models.
The output is a set of synthesized scenes which support the activities of the captured real environments.

\begin{figure}[!t] \centering
    \includegraphics[width=\linewidth]{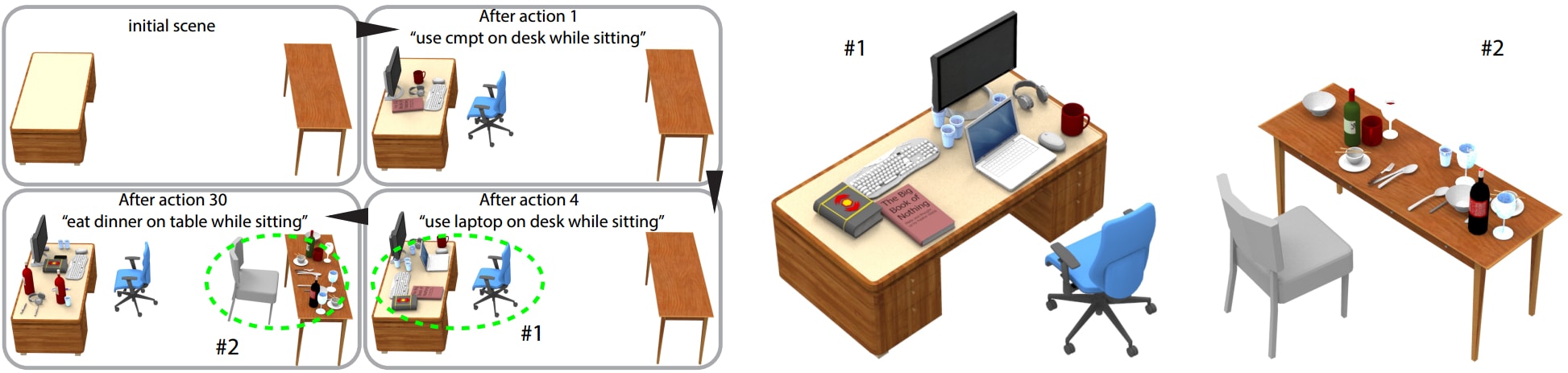}
    \caption{Action-driven scene evolution~\cite{ma2016} alters an initial scene consisting of a desk and a dining table. The initial scene and three intermediate evolution snapshots (after 1, 4, and 30 actions, respectively) are shown (left) with zoom-ins for better viewing on the right.}
    \label{fig:ma2016}
\end{figure}
Real-world scenes are not static environments, instead they evolve over time, driven by object movements resulting from human.
Ma et al.~\cite{ma2016} introduce a framework for \emph{action-driven evolution} of 3D indoor scenes, where the goal is to simulate how scenes are progressively altered by human actions (Figure~\ref{fig:ma2016}).
In this work, an action is a typical activity performed by humans in an indoor environment and involves one or more human poses, objects and their spatial configurations.
These pieces of information are learned from annotated photographs~\cite{lin2014} and summarized into a set of \emph{action models}.
\emph{Correlation} between the learned actions are analyzed to guide the construction of an \emph{action graph}, whose nodes correspond to actions and edges encode correlations between actions in the form of transitional probabilities.
Sequences of actions are sampled from the action graph, where each action triggers appropriate object placements including relocating existing objects or inserting new objects related to the action, leading to a continuous and progressive scene evolution.
After applying a few actions to the initial scene, realistic and messy 3D scenes can be generated.

	\section{Conclusion and Open Problems}

This report surveys recent research progress on geometry, structure and semantic analysis of 3D indoor data and different modeling techniques for creating plausible and realistic indoor scenes.
Data-driven approaches which utilize existing large-scale scene data have produced promising results in scene understanding and modeling.
However, the goal of generating truly realistic and high-quality 3D indoor scenes that are readily usable for applications such as 3D games is still not achieved.
More efforts from related research communities, e.g., computer graphics, vision and robotics, are needed to solve the various remaining challenges in scene modeling.
Here, three open problems that may bring some interests and inspire future works for indoor scene processing are listed as follows.

The key for data-driven approaches is the source of data, especially those with rich and semantic annotations.
Recent works in computer vision have developed easy-to-use scene capture and annotation tools.
Large-scale richly-annotated scene databases such as SceneNN~\cite{hua2016} and ScanNet~\cite{dai2017}, which contain hundreds to thousands of scenes captured from real-world environments, are now available.
It remains a problem that how to utilize such data to improve the generation of realistic and detailed 3D scenes.

Natural language is arguably the most accessible input for content creation.
To date, most text-to-scene systems~\cite{chang2014b, coyne2012, seversky2006} focus on mapping explicit scene arrangement languages to placements of individual objects.
Synthesizing complex scenes by explicitly specifying each object is tedious due to the number of objects and the inherent ambiguity of language.
A more efficient text-driven scene synthesis system is needed for creating complex 3D scenes using compact language commands.

Existing scene synthesis methods have taken account of spatial relationships~\cite{fisher2012, zhao2016}, functionality or action constraints~\cite{fisher2015, ma2016} among objects during the scene generation.
The style compatibility among objects which is also important for plausibility of scenes are not considered in the scene synthesis task.
Current research for style analysis mainly work on learning a style metric~\cite{liu2015b, lun2015}, identifying style-defining elements~\cite{hu2017}, or synthesis of 3D model geometry with a given style.
Introducing style compatibility, e.g., geometry style of objects and layout style of scenes, to synthesis of complex scenes is a promising future direction.

	\bibliographystyle{plain}
	\bibliography{references}

\end{document}